\documentclass[a4paper,fleqn,usenatbib]{mnras}

\usepackage[T1]{fontenc}
\usepackage{ae,aecompl}

\usepackage{graphicx}	
\usepackage{amsmath}	
\usepackage{amssymb}	

\bibpunct{(}{)}{;}{a}{}{,}

\newcommand{\ks}{\hbox{$K_{\rm s}$}}

\title[The old open cluster LS\,94]{GeMs/GSAOI observations of La Serena\,94: an old and far open cluster inside the solar circle\thanks{Based on observations obtained at the Gemini Observatory, which is operated by the Association of Universities for Research in Astronomy, Inc., under a cooperative agreement with the NSF on behalf of the Gemini partnership: the National Science Foundation (United States), the Science and Technology Facilities Council (United Kingdom), the National Research Council (Canada), CONICYT (Chile), the Australian Research Council (Australia), Minist\'erio da Ci\^encia, Tecnologia e Inova\c c\~ao (Brazil) and Ministerio de Ciencia, Tecnolog\'ia e Innovaci\'on Productiva (Argentina).}}

\author[J. F. C. Santos Jr. et al.]{
Jo\~ao F. C. Santos Jr.$^1$\thanks{E-mail: jsantos@fisica.ufmg.br}, 
Alexandre Roman--Lopes$^2$,
Eleazar R. Carrasco$^3$,
\newauthor{Francisco F. S. Maia$^{4}$, Benoit Neichel$^{5}$} 
\\
$^{1}$Departamento de F\'isica, ICEx, Universidade Federal de Minas Gerais, Av. Ant\^{o}nio Carlos 6627, 31270-901 Belo Horizonte,\\ MG,
Brazil\\
$^2$Departamento de F\'isica y Astronom\'ia, Universidad de la Serena, Av. Juan Cisternas 1200 Norte, La Serena, Chile\\
$^3$Gemini Observatory/AURA, Casilla 603, La Serena, Chile\\
$^4$Institut de Plan\'etologie et d'Astrophysique de Grenoble (IPAG) UMR 5274,
Grenoble, F-38041, France\\
$^5$Aix Marseille Universit\'e, CNRS, LAM (Laboratoire d'Astrophysique de Marseille) UMR 7326, 13388 Marseille, France
}

\date{Accepted 18 November 2015}

\pubyear{2015}

\begin{document}
\label{firstpage}
\pagerange{\pageref{firstpage}--\pageref{lastpage}}
\maketitle

\begin{abstract}
Physical properties were derived for the candidate open cluster La Serena\,94,
recently unveiled by the VVV collaboration. Thanks to the exquisite angular 
resolution provided by GeMS/GSAOI, we could characterize this system in 
detail, for the first time, with deep photometry in $JHK_{\rm s}$--bands.
Decontaminated $JHK_{\rm s}$ diagrams reach  
about 5 mag below the cluster turnoff in $H$. The locus of red clump
giants in the colour--colour diagram, together with an extinction law, 
was used to obtain an average extinction of $A_V=14.18\pm0.71$. The same
stars were considered as standard--candles to derive the cluster distance,
$8.5\pm 1.0$\,kpc. Isochrones were matched to the cluster 
colour--magnitude diagrams to determine its age, 
$\log{t({\rm yr})}=9.12\pm 0.06$, and metallicity, $Z=0.02\pm0.01$.
A core radius of $r_{\rm c}=0.51\pm 0.04$\,pc was found by fitting
King models to the radial density profile.
By adding up the visible stellar mass to an
extrapolated mass function, the cluster mass was 
estimated as $\mathcal{M}=(2.65\pm0.57)\times 10^3$\,M$_\odot$, 
consistent with an integrated magnitude of $M_{K_{\rm s}}=-5.82\pm0.16$ 
and a tidal radius of  $r_{\rm t}=17.2\pm2.1$\,pc.
The overall characteristics of La Serena\,94 
confirm that it is an old open cluster located in the Crux spiral arm
towards the fourth Galactic quadrant and distant $7.30\pm 0.49$\,kpc from
the Galactic centre. The cluster distorted structure, mass segregation
and age indicate that it is a dynamically evolved stellar system.

\end{abstract}




\section{Introduction}

Galactic open clusters are key to the development of the theories on formation and evolution of galaxies.
Indeed, in the last decades the studies of Galactic clusters have proven to be extremely important astrophysical laboratories for a wide range of problematic issues, particularly those pertaining to the disc abundance gradients and
age--metallicity relations \citep{pca95,cnp98,hcc02,mrd15}.
The knowledge and accurate measurement of clusters' fundamental parameters like age, heliocentric distance, 
reddening, metallicity, mass and size, play a key role in studies of the Milky Way (MW) global properties, such as its formation history \citep{b1} 
and dynamical properties \citep{dl05}.
In this sense, the study of the stellar populations of old open clusters may contribute to answer some fundamental questions related to the structure and 
evolution of the Galaxy during its early formation time 
\citep[][and references therein]{b2,b3,b4}. 

In one hand, the study of old open clusters in the Galactic plane, 
particularly those situated towards the first and fourth quadrants 
inside the solar circle, is problematic due to the high and 
patchy extinction, which makes
optical observations difficult to impossible along most lines of sight. Also, crowding may be a limitation, mainly in directions where
more than one spiral arm may be present. With the advent of deep near--infrared surveys like 2MASS \citep{b13}, VVV \citep{mle10}, and WISE \citep{b14},
hundreds of new cluster candidates have been found in the past years 
\citep{b5,b6,b7,b8,brn15,b10}, making possible further 
study of {\it new} old open cluster candidates 
placed inside or near the solar circle \citep{b1,b11}. 
On the other hand, open clusters older than $\sim$1\,Gyr
are normally found near the solar circle and/or in the
outer Galaxy \citep{b1}, where dynamical interactions with giant 
molecular clouds and the disk is less common \citep{b12}.

Adaptive optics systems are especially useful to the investigation of relatively
compact, obscured and distant star clusters in the Galactic disc \citep{mob08}.
We present a study of La Serena\,94
(hereafter LS\,94), localized in the fourth Galactic quadrant.
The object is one of the star cluster candidates detected in 
the VISTA Variables in the V\'{\i}a L\'actea ESO public 
survey  by \citet{brn15}. 
The present study is based on high spatial resolution, near--infrared images
obtained with the first Multi--Conjugate Adaptive Optics system in use 
in a 8--m telescope.

The paper is organized as follows. In Section~\ref{sec:obs} we describe 
the observations and 
data reduction, including the point spread function analysis and the 
photometric calibration of the stars
detected in the observed field. The centre 
determination and the stellar density map of the cluster candidate are 
presented in Section~\ref{sec:map}. Section~\ref{sec:stpop} contains a 
detailed analysis of LS\,94 stellar population concerning its radial 
variation and the determination of photometric membership from decontaminated 
photometry. In Section~\ref{sec:red_dist}, the cluster fundamental parameters 
reddening, distance, age and metallicity are derived. An investigation of the 
structural properties and their consequences is presented in 
Section~\ref{sec:struc}. Section~\ref{sec:lm} shows the 
luminosity and mass functions, 
built to derive the cluster overall luminosity and mass, from which the 
tidal radius is estimated.
The results are discussed in Section~\ref{sec:dis} and the conclusions
given in Section~\ref{sec:conc}.

\section{Data acquisition and reductions}
\label{sec:obs}

\subsection{Observations}

The observations of LS\,94 were obtained with the Gemini--South telescope using 
the Gemini South Adaptive Optics Imager \citep[GSAOI --][]{mhs04,cem12} and
the Gemini South Multi--Conjugate Adaptive Optics System \citep[GeMS --][]{rnb14,nrv14}.
GeMS is a facility Adaptive Optics (AO) system for the Gemini South telescope. This AO system 
uses five sodium Laser Guide Stars (LGSs) to correct for atmospheric distortion and up to three 
Natural Guide Stars (NGSs) brighter than $R = 15.5$ mag to compensate for tip--tilt and plate 
modes variation over a 2 arcmin ~field--of--view (FoV) of the AO bench unit \citep[CANOPUS,][]{rnb14}. 
GSAOI is a near--infrared AO camera used with GeMS. Together, the two facility instruments
can deliver near--diffraction limited images in the wavelength interval of 
0.9 -- 2.4$\micron$. 
The GSAOI detector is formed by $2 \times 2$ mosaic Rockwell HAWAII-2RG $2048 \times 2048$ 
arrays. At the f/32 GeMS output focus, GSAOI provides a FoV of 85~$\times$ 
85~arcsec$^2$~on the sky with a 0.02 arcsec per pixel sampling and gaps of $\sim 3$\,arcsec between arrays.

The cluster LS\,94 was imaged through the $J$ (1.250 $\micron$), $H$ (1.635 $\micron$) and 
$K$~(2.200 $\micron$) filters during the night of May 22 -- 23, 2013, as part of the program 
GS-2012B-SV-499 (GeMS/GSAOI commissioning data
\footnote{\tiny\url{http://www.cadc-ccda.hia-iha.nrc-cnrc.gc.ca/en/gsa/sv/dataSVGSAOI_v1.html}}). 
For each filter, 9 images of 60 seconds were obtained, providing an effective
exposure time of 540 seconds. An offset of 4 arcsec  between individual images, following a
$3 \times 3$ dither pattern, was used to fill the gaps between arrays. Because LS\,94 is 
located in a crowded sky region in the Galactic plane, sky frames were observed in a separate
region located 10 degrees North of the position of the cluster centre using the same
dither pattern and offset size as the main target. The images were obtained under photometric 
conditions and variable seeing. The values for the natural seeing from the DIMM monitor at Cerro Pach\'on, the 
average resolution derived from stars over the field per filter (AO FWHM, see Sec.~\ref{photo})  and 
the average Strehl ratios are shown in Table \ref{tab:obslog}.  

\begin{table}
\tiny
\centering
\caption{Observing log}
	\label{tab:obslog}
	\begin{tabular}{lccccc}
		\hline
		Filter & Exp. Time & Airmass & Seeing  & AO FWHM & Strehl \\  
		      & [s]    &         & [arcsec]& [mas] & Ratio [\%]\\ 
		\hline
		$K$    & $9 \times 60$ & 1.202 & 0.79$\pm$0.10 & 98$\pm$13 & $12\pm2$ \\
		$H$    & $9 \times 60$ & 1.212 & 0.84$\pm$0.12 & 102$\pm$15 & $9\pm2$  \\
		$J$    & $9 \times 60$ & 1.223 & 1.03$\pm$0.22 & 211$\pm$22 & $3\pm2$ \\
		\hline
	\end{tabular}
\end{table}

\subsection{Data reduction}

The data were reduced following the standard procedures for near--infrared imaging provided
by the Gemini/GSAOI package inside \textit{IRAF} \citep{t86}. Each science image was processed with the program \textit{GAREDUCE}.
The arrays in the science frames were corrected for non--linearity, 
subtracted off the sky, divided  
by the master domeflat fields image, and multiplied by the GAIN to convert from ADU to electrons. 

Prior to mosaic the science frames and create images with a single extension, it is necessary to
remove the instrumental distortion produced by the off-axis parabolic system used in GeMS \citep{rnb12,rnb14}.
The instrumental distortion was corrected using a high--order distortion map derived from an astrometric 
field located in the Large Magellanic Cloud.  The positions of the stars in this field were derived from the 
HST/ACS data (HST Proposal 10753, PI: Rosa Diaz-Miller, Cycle 14). The distortion map was derived using 
the position of about 300 stars uniformly distributed across the GSAOI detector. The distortion map has a 
star--position accuracy less than $\sim$ 0.1 arcsec.  We used the program \textit{MSCSETWCS} inside
the \textit{MSCRED} package to apply the distortion correction to each GSAOI array. The program
\textit{MSCIMAGE} was employed to resample each GSAOI multi--extension frame into a single image 
and to a common reference position.

Unfortunately, the distortion correction applied above does not remove all the instrumental distortion. 
There is a dynamic distortion component \citep[see][]{rnb12,nlr14} which depend 
on the location of the NGSs in the GeMS patrol field, the size of the offsets and the dither pattern used. 
This effect is more pronounced in the outer parts of the mosaic--ed images where the position of a given 
star in the different images can have variations of up to 10 pixels. Moreover, given the spatial resolution 
of our GeMS/GSAOI images, in particular for the $K$--band (see Table \ref{tab:obslog}), this effect has 
to be corrected, otherwise the co--addition will be wrong. To correct the dynamic distortion and co--add the 
images, we have used a modified version of the \textit{IMCOADD} program inside the GEMINI/GEMTOOLS 
package. For each filter, the first image is used as a reference to search for stars using the \textit{NOAO/DAOFIND} 
program. Then, a geometrical transformation is derived  to register the images to a common pixel position using 
unsaturated stars in common between the images with the \textit{GEOMAP} program. The derived transformation 
is applied to each image with the \textit{GEOTRANS} program. Lastly, the images are combined by averaging the 
good pixels. The \textit{rms} of the resulting fit in the individual images was less than 0.1 pixels.

\begin{figure*}
\includegraphics[width=0.9\textwidth]{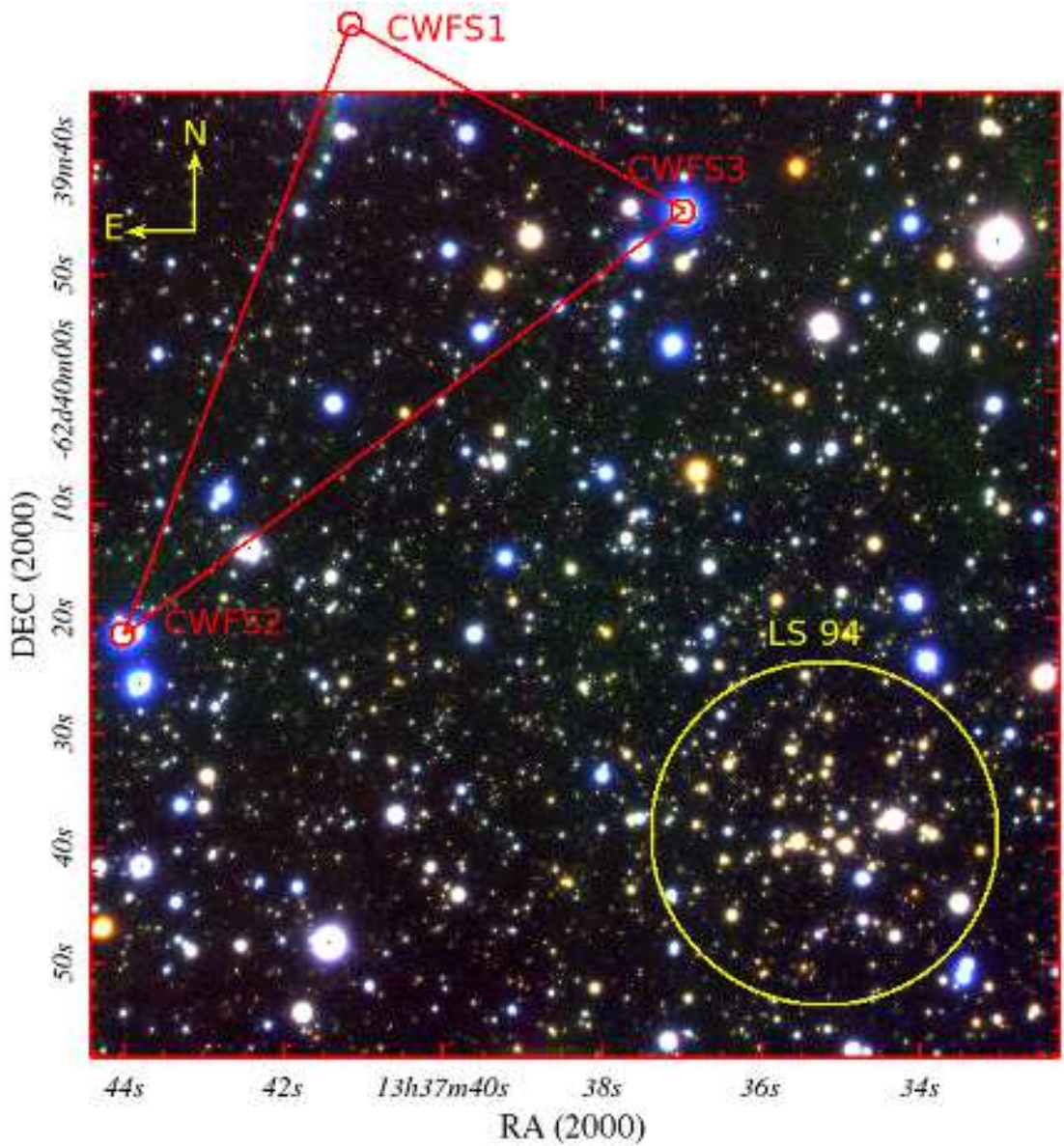}
\caption{ $J$ (blue), $H$ (green), $K$ (red) colour composite image 
of LS\,94. The yellow circle (30\,arcsec of diameter) 
indicates the position of the cluster candidate. The red small circles 
show the location of the natural guide stars used during 
the observations. The size of the image is 85$\times$85\,arcsec$^2$.}
\label{fig:ls94_color}
\end{figure*}

The WCS in the final co--added images need to be calibrated in order to have all filters registered to a common
WCS and pixel positions. The WCS was calibrated using a catalogue of non--saturated stars
derived from the VIRCAM/VVV \ks~image and uniformly distributed across the GSAOI FoV. We used the
\textit{CCMAP} program to derive a linear transformation (translation, scale and rotation) to correct the 
WCS in all co--added images. The final co--added images have an accuracy in the WCS solution (average 
astrometric error) of $\sim$ 0.05 arcsec. Fig. \ref{fig:ls94_color} shows the $JHK$ colour composite 
image of LS\,94. The big circle indicates the location of the cluster candidate. The positions of
the NGSs are depicted. The corrected AO FWHM and Strehl ratios derived from the final co--added images 
are presented in Table \ref{tab:obslog}.

\subsection{Photometry}
\label{photo}

\begin{figure}
                   \includegraphics[width=1.0\linewidth]{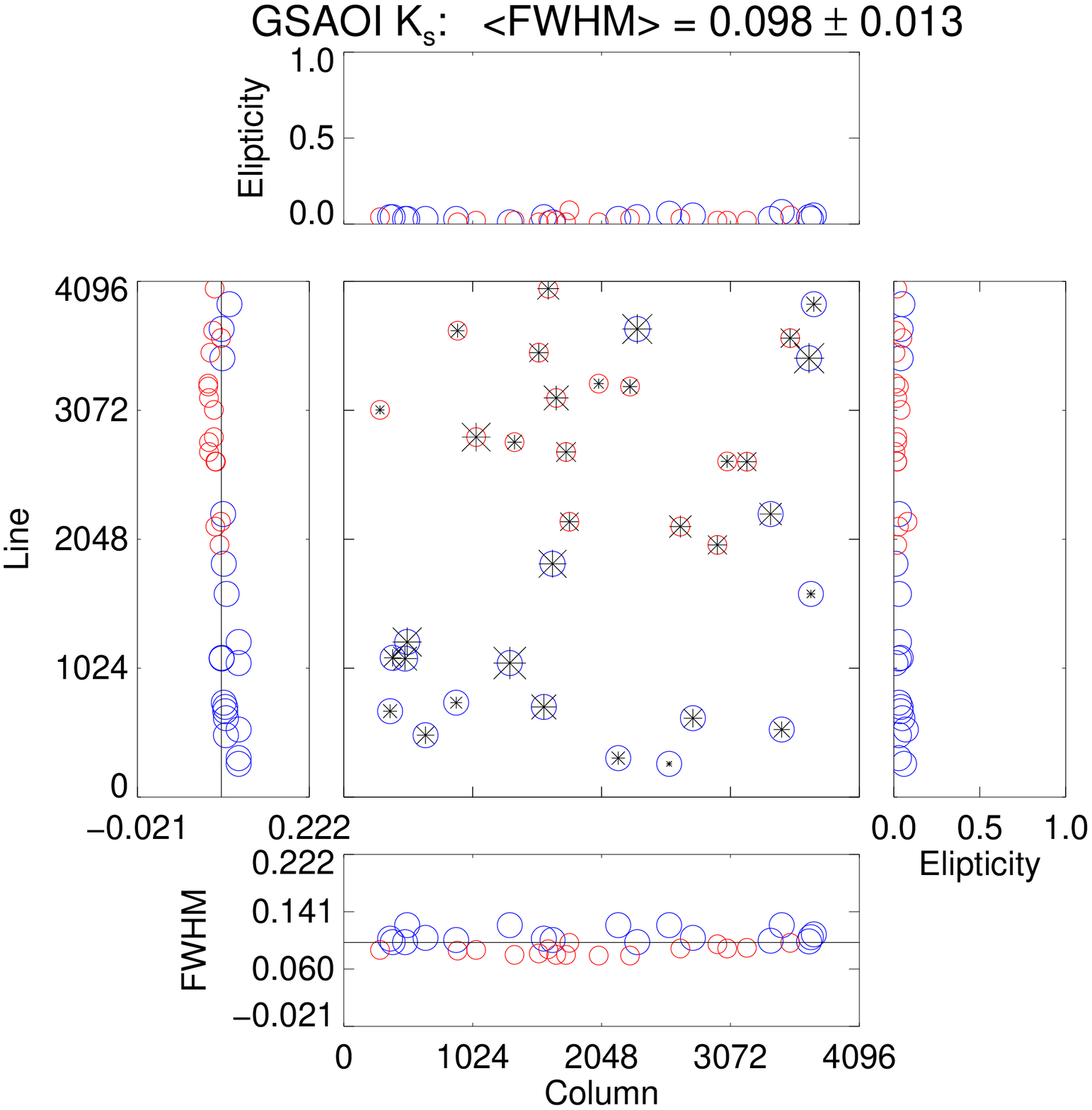}
\caption{Spatial variation of the PSF for the GSAOI $K$ combined image,
with the FWHM values given in arcsec. Asterisk sizes indicate the magnitude
of the stars
while circle sizes represent the FWHM of the stars, which is blue 
if above the average and red if below the average.}
\label{fig:isop}
\end{figure}

The program \textit{starfinder} \citep{dbb00} was used to perform point spread function (PSF) 
photometry. The PSF model was built from several (typically 40) 
relatively bright, 
isolated stars uniformly distributed across the frames. 
Isoplanatism was evaluated for the pre--reduced, combined images using 
\textit{IRAF} task \textit{psfmeasure}. 
Fig.~\ref{fig:isop} gives information on the
spatial variation of the delivered image quality of randomly located 
stars in the $K$ frame.
These stars were used to built the average PSF employed in the photometric 
reduction. Fig.~\ref{fig:isop} shows also how the stellar profile parameters
FWHM and ellipticity vary along lines and columns of the detector. 
The circle sizes
indicate the observed FWHM (blue if it is above the average and red if it is 
below the average), and the asterisk sizes indicate the relative magnitude
of the stars. There is an evident tendency
for smaller FWHM and ellipticities to lie in the superior part of 
the frame, where the NGS are located and, in consequence, 
the AO correction performed better. However,
the difference towards opposite sides of the frame are small, as
reflected by the dispersion of the FWHM, $0.098\pm0.013$\,arcsec.

The same analysis was performed for the $J$ and $H$ combined frames. 
Although there is a degradation of the image quality compared to the 
$K$ band, as expected for shorter wavelengths, it is minor. 
The anisoplanatism  is even less noticeable for $J$ and $H$ 
than for $K$. The FWHM is $0.211\pm0.022$\,arcsec and 
$0.102\pm0.015$\,arcsec for  $J$ and $H$, respectively. 

The extinction coefficients\footnote{\url{https://www.gemini.edu/node/10781?q=node/10790}} ($k_J=0.015$, $k_H=0.015$, $k_K=0.033$) and an 
initial zeropoint (25.0 mag) were adopted to transform the PSF flux into
instrumental magnitudes, which were calibrated to the 2MASS photometric system.
With this aim, we resort to the VISTA Variables in the V\'ia L\'actea 
(VVV) survey 
\citep{mle10,sha10}, which collected near--infrared photometry of selected regions of our Galaxy disc 
and bulge with the VISTA 4--m telescope (Visible and Infrared Survey Telescope for Astronomy). Specifically, the positions of stars in the VVV photometry 
in common with the GSAOI FoV were matched
and the instrumental magnitudes from GSAOI calibrated against VVV magnitudes
(in the 2MASS photometric system).

\begin{figure}
	\includegraphics[width=\columnwidth]{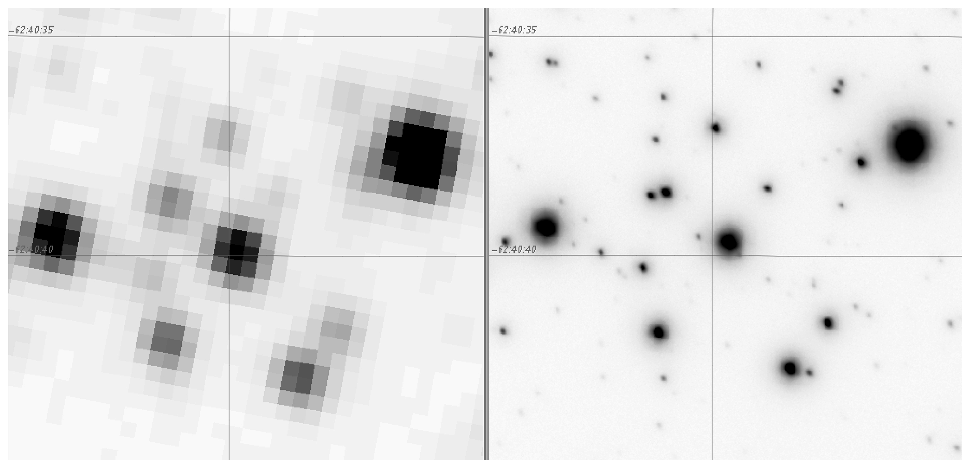}
\caption{VVV $K_{\rm s}$ (left) and GSAOI $K$ (right)  images are compared for the cluster core region. Each frame is $12\times 13$\,arcsec$^2$. North is to the top and East to the left.}
    \label{fig:image}
\end{figure}

\begin{figure}
	\includegraphics[width=\columnwidth]{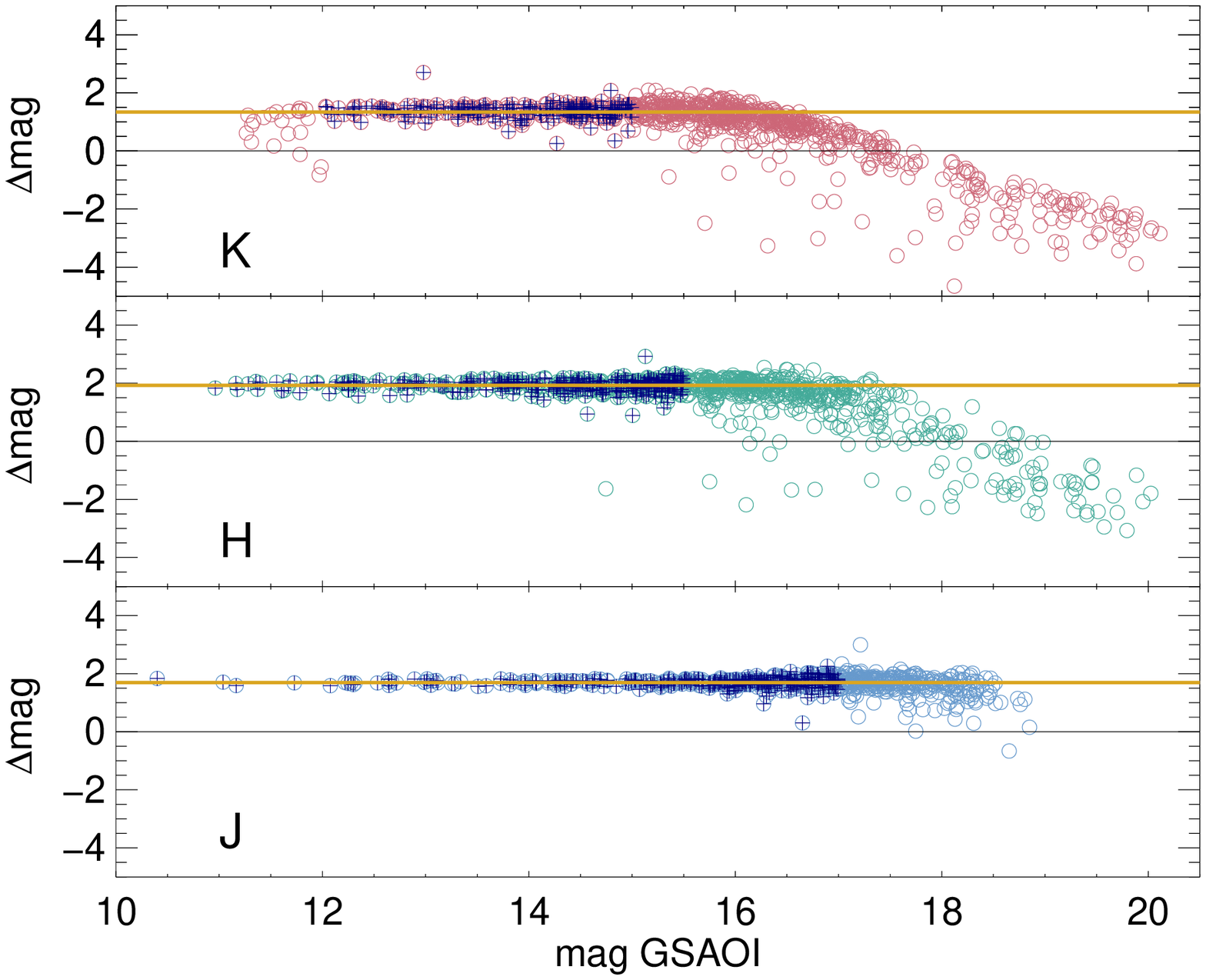}
    \caption{The difference between the VVV magnitude and the GSAOI (instrumental) magnitudes. A constant 
(straight line) was fit to selected ranges of magnitude (plusses) 
excluding stars affected by saturation, unresolved binaries and 
crowding. }
    \label{fig:calib}
\end{figure}

Fig.~\ref{fig:image} contrasts VVV $K_{\rm s}$ and GSAOI $K$ images of the central regions
of the cluster evidencing the improvement of spatial resolution and photometric 
deepness of GSAOI over VVV.
The difference between the VVV magnitude and
the GSAOI (instrumental) magnitude for $JH$ filters is 
shown in Fig.~\ref{fig:calib}. The same difference 
is shown for the $K$ band, but it refers to the 2MASS $K_{\rm s}$ 
(2.150\,$\micron$) in the case of VVV magnitudes. For all bands, 
the data distribution could be fit by a zero-order polynomium (constant),
except where stars are saturated (only in the $K$--band) or 
affected by unresolved binaries and crowding in VVV data.
Therefore, a constant was fit
to selected magnitude ranges excluding these stars (darker symbols in 
Fig.~\ref{fig:calib}).
The straight line represents the average weighted by the magnitude 
uncertainties, mathematically:

\begin{equation}
J(VVV)-J(GSAOI)=1.69530 \pm 0.00089
\end{equation}
\begin{equation}
H(VVV)-H(GSAOI)=1.92705\pm0.00070
\end{equation}
\begin{equation}
K_{\rm s}(VVV)-K(GSAOI)=1.3393 \pm0.0015
\end{equation}

\begin{figure}
	\includegraphics[width=\columnwidth]{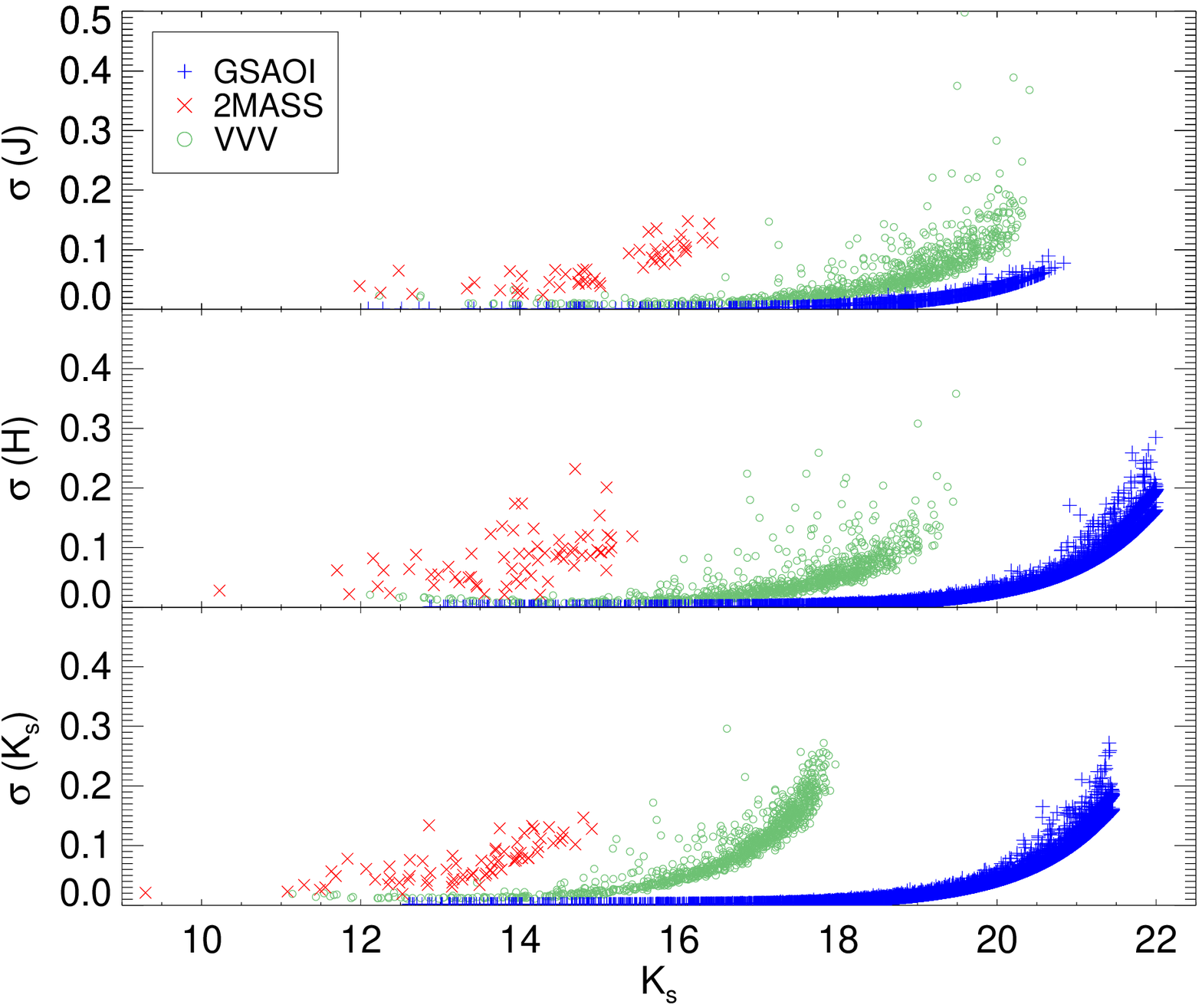}
\caption{Uncertainties in $J$, $H$ and $K_{\rm s}$ for 2MASS (red crosses), 
VVV (green circles) and GSAOI (blue plusses) data compared.}
    \label{fig:errors}
\end{figure}

\begin{figure}
	\includegraphics[width=\columnwidth]{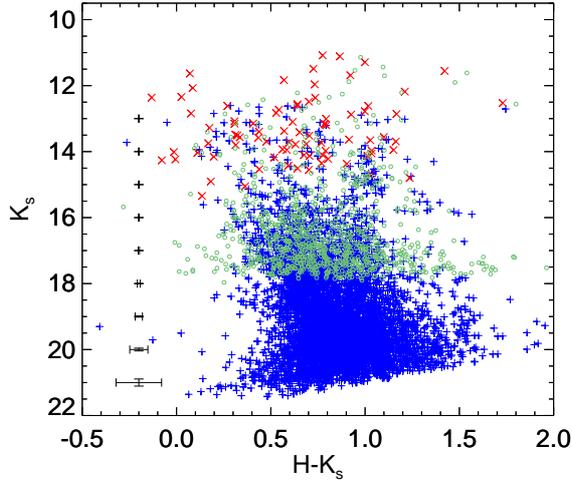}
\caption{Colour--magnitude diagram with data from 2MASS, VVV and GSAOI in the
$85\times 85$\,arcsec$^2$ FoV of GSAOI. Uncertainties are indicated 
by error bars on the left. The symbols are the same as in 
Fig.~\ref{fig:errors}.}
    \label{fig:cmd_HK_K}
\end{figure}

\begin{figure}
\centering
	\includegraphics[width=0.8\columnwidth]{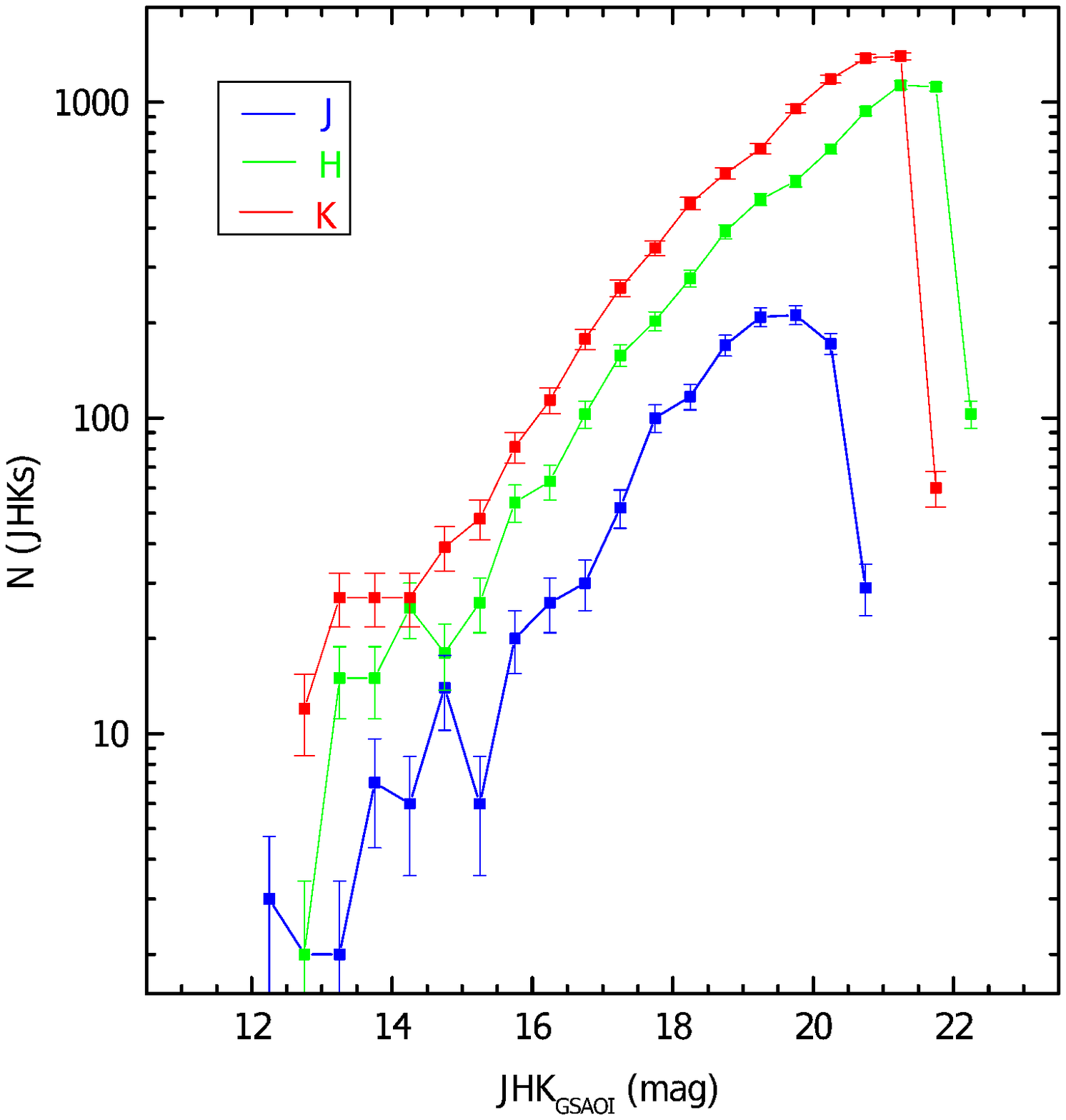}
\caption{Number of stars detected as a function of magnitude for each 
near--infrared band.}
    \label{fig:compl}
\end{figure}

The final magnitude uncertainties were obtained by propagation considering
the above calibration errors and the PSF errors. Magnitudes of GSAOI 
saturated stars were replaced by 2MASS magnitudes. This procedure only affects 
stars with $K < 12.0$.  
Fig.~\ref{fig:errors} shows the final magnitude errors of GSAOI data 
compared to data from 2MASS and VVV in the field of GSAOI. The same is
presented in Fig.~\ref{fig:cmd_HK_K} for the CMD 
$K_{\rm s}$ vs $H-K_{\rm s}$, where the uncertainties of GSAOI photometry
are indicated by error bars.  Along the Galactic disc, we expect a 
rising number of stars as the magnitude increases, until source 
confusion associated to the instrumental sensitivity cause this number to
drop making the data no longer complete. Fig.~\ref{fig:compl} 
shows this trend, where the magnitude for which the star counts peak 
indicates the estimated $\sim 100\%$ completeness limits of 19.3,
21.2 and 20.7 magnitudes for $J$, $H$ and 
$K_{\rm s}$, respectively.

\section{Cluster centre and stellar density map}
\label{sec:map}

\begin{figure}
	\includegraphics[width=\columnwidth]{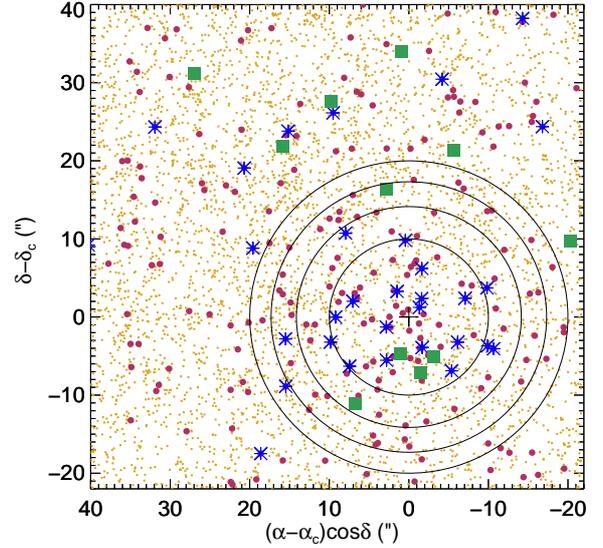}
\caption{ Stellar density map of LS\,94 and field in the GSAOI FoV. 
(a)~map limited in colour ($H-K_{\rm s} > 0.8$) and magnitude ($K_{\rm s} < 18$) to enhance the cluster contrast against the field. The centre (cross) and the adopted radius to search for this centre (circle) are indicated. (b)~map complementary to that in (a), i.e., data selected for $H-K_{\rm s} < 0.8$ and $K_{\rm s} < 18$, to exclude cluster stars.} 
    \label{fig:dens}
\end{figure}

The equatorial coordinates of the candidate cluster were catalogued 
by \citet{brn15} 
as $\alpha_{\rm J2000}=13^{\rm h}37^{\rm m}35.02^{\rm s}$ and 
$\delta_{\rm J2000}=-62\degr40\arcmin36.8\arcsec$,
with  Galactic coordinates $l=308\fdg199$ and $b=-0\fdg278$.
The object centre coordinates were redetermined with the GSAOI data.
The photometry was filtered to enhance the contrast between cluster and 
field stars: data in the range $K_s < 18$ and $H-K_s >0.8$ yields the
stellar density map shown in  Fig.~\ref{fig:dens}a. Such cutoffs select
most of the cluster stars (see Sect. \ref{sec:decont}). 
On the other hand, data in the range $K_s < 18$ and $H-K_s <0.8$ emphasizes 
the density map for field stars in Fig.~\ref{fig:dens}b.
As can be seen in Fig.~\ref{fig:dens}, the cluster core presents 
an elongation nearly along the equatorial 
N-S direction, roughly aligned with Galactic N-S.

The centre determination relies on an algorithm that averages the 
stars' coordinates within a circle of radius 12\,arcsec, approximately 
the object core radius (see Sect.~\ref{sec:struc}). This 
centre realocates
the circle and a new centre is calculated yelding a new shift of the circle. 
The process is repeated until the difference between consecutive 
centres are smaller than a previously set quantity. To obtain a more physically 
meaningfull quantity, the density--weigthed centre was calculated using
the same algorithm, but considering the stellar density at each star
position as weight. The final equatorial coordinates derived from the procedure
described above are: 
 $\alpha_{\rm J2000}=13^{\rm h}37^{\rm m}35.4^{\rm s}$ and 
$\delta_{\rm J2000}=-62\degr40\arcmin34.7\arcsec$.

The cluster calculated centre and core radius  
are indicated by a cross and a circle, respectively, in Fig.~\ref{fig:dens}. 
Note the object asymmetry revealing a 
disturbed stellar distribution and/or variable extinction.
Since the Galactic longitude at this position runs almost parallel to the
right ascension, may be we are witnessing the disruption of the cluster 
as a consequence of its interaction with the Galactic disc tidal 
field. But also its appearance could be partially explained 
as an artefact of differential reddening produced by filamentar 
interestelar clouds in front of the system. Spitzer images revealing
the dust distribution 
in the region, on the contrary, do not show any particular 
feature which could possibly obscure stars preferentially in any direction 
around the cluster location.
The relatively small FoV of GSAOI ($85\times 85$\,arcsec$^2$), is compensated by its excelent spatial resolution, which makes evident the cluster possible distorted 
morphology. This point is further discussed in 
Sect.~\ref{sec:red_dist}.

\section{Analysis of the stellar population}
\label{sec:stpop}

\subsection{Radial variations}

\begin{figure}
	\includegraphics[width=\columnwidth]{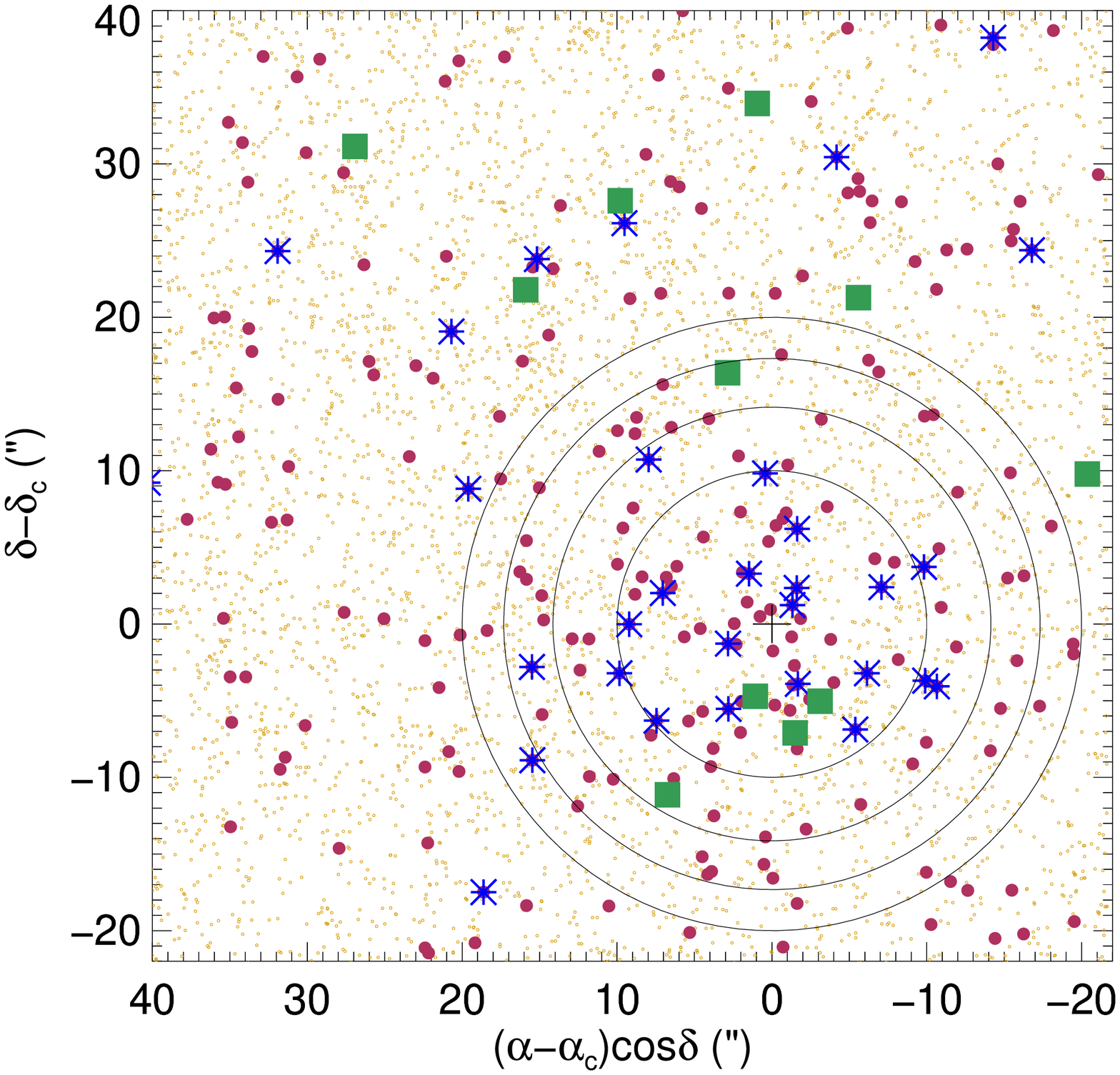}
\caption{Spatial stellar distribution in the cluster region. Different symbols identify clump giant stars (RCG, blue asterisks), giant branch stars (green squares), 
main sequence stars (MS, red dots), field stars (small yellow dots). The black circles centred in the cluster delimit regions of identical area.}
    \label{fig:fields}
\end{figure}

\begin{figure}
	\includegraphics[width=\columnwidth]{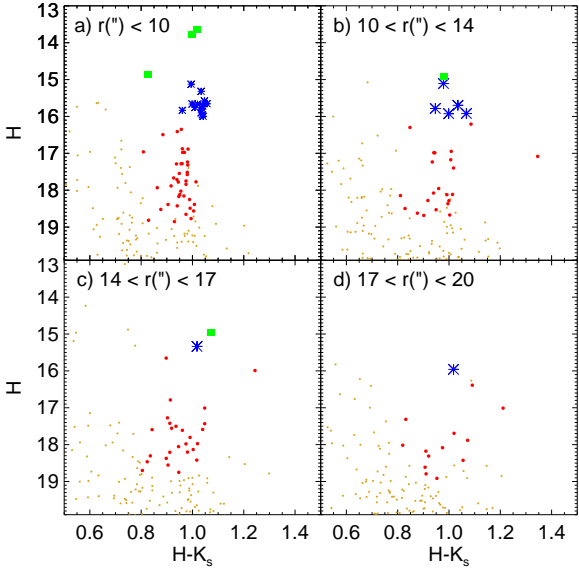}
\caption{CMD $H$ vs $H-K_{\rm s}$ for each annular region shown in 
Fig.~\ref{fig:fields}. The red dots are limited in colour 
($H-K_{\rm s} > 0.8$) and magnitude ($K_{\rm s} < 18$).
Symbols as in Fig.~\ref{fig:fields}.}
    \label{fig:HK_H}
\end{figure}

\begin{figure}
	\includegraphics[width=\columnwidth]{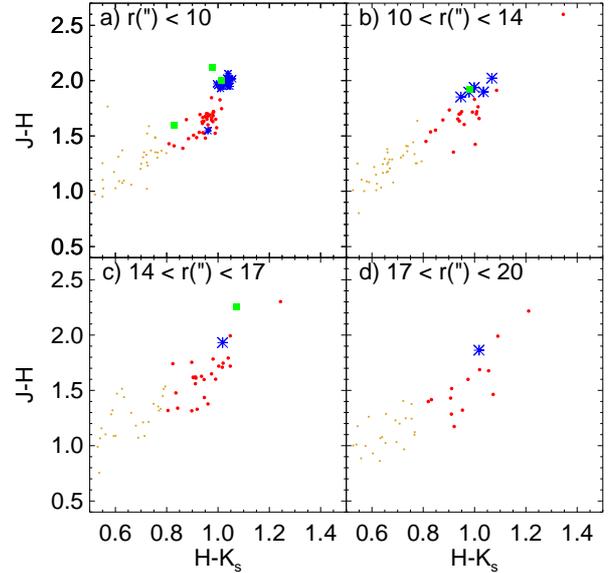}
\caption{Same as Fig.~\ref{fig:HK_H} but for $J-H$ vs $H-K_{\rm s}$.}
    \label{fig:HK_JH}
\end{figure}

The analysis of the stellar content in annular regions of same area
(Fig.~\ref{fig:fields}) is presented in Figs.~\ref{fig:HK_H} and 
\ref{fig:HK_JH}, which show photometric diagrams evidencing the progressive 
changes in 
the number of stars in different evolutionary stages from the cluster 
centre to the periphery. The photometry was filtered to show stars with 
$H-K_{\rm s} > 0.8$ and $K_{\rm s} < 18$. The inner circular field has 
radius 10\,arcsec
and the three subsequent external annuli, of same area as the inner 
circle, are bounded by 14, 17 and 20\,arcsec. 
 
There is a tendency of red clump giant (RCG) stars to group by the 
cluster very centre, with most of them confined within 14\,arcsec.
The same is true for the red giants, but there are few of them, therefore
this tendency can not be assured. Although main sequence (MS) stars
are also concentrated within $r<10$\,arcsec, there are many of them 
throughout the region up to 20\,arcsec. To a deeper analysis of the
stellar population it is necessary to disentangle the cluster members from
the stars in the general Galactic disc field, which is done in the next 
Section.

\subsection{CMD decontamination method and photometric membership}
\label{sec:decont}

To disentangle cluster member stars from the contaminating stellar field it
was employed a method that has been developed, tested and applied to 
Galactic open clusters. It recovers statistically the cluster 
intrinsic stellar population assigning
membership probabilities to each star \citep{mcs10}. 

\subsubsection{The method}

The decontamination method deals with stellar photometry of the cluster 
and adjacent fields, both of same area. A CMD 
is built for both field and cluster plus field and divided in retangular cells
of sizes corresponding roughly to ten times the average uncertainties 
in magnitude and colour. Because the GSAOI data is deeper for $H$ and 
$K_{\rm s}$ than for $J$, the former magnitudes were used in the 
decontamination procedure.
After the initial setup, the number of stars is counted for the
cluster region ($N_{clu+field}$) and for the control field ($N_{field}$)
for every corresponding cell in both CMDs. A preliminary decontaminated 
sample is generated by removing the expected number of field stars from the 
cluster+field cells, prioritising the exclusion of the stars farther 
away from the cluster centre.  
An initial membership probability was assigned to all stars
in the cluster region (even those removed from the CMD) according to 
their overdensity in each CMD cell 
relative to that in the field, i.e., $P=(N_{clu+field}-N_{field})/N_{clu+field}$.
For cells containing more field stars than cluster stars, a zero probability
was adopted.

To minimize the sensitivity of the method to the choice of initial parameters,
the procedure is repeated for different sizes and positions of the
cells. Their sizes are compressed and expanded by one third from the initial
value (in both mag and colour) and their positions are shifted also by one
third of the initial amounts towards negative and positive values. In total,
729 grid configurations are employed and the decontamination procedure
described above is performed for each of them. An exclusion index is then 
defined
as the number of times in which a given star is removed from the CMD. The
final decontaminated sample is built by removing stars from the CMD with 
exclusion index above a predefined threshold. Similarly, the final photometric
membership probability is obtained from the average of the membership 
probabilities assigned to each star.

Both the membership probability and the exclusion index actuate as
complementary indicators since field stars can be identified for their low
membership probability as well as for their high exclusion index.
Tests of the method applied to photometry of Galactic open clusters and 
simulations of simple stellar populations indicated that a succesful
decontamination is obtained when the exclusion index is around 80\% (stars
that are excluded in more than 80\% of the 729 grid configurations are removed
from the CMD) and the sample retains stars with membership probability 
above 30\%. These thresholds were adopted  for LS\,94.
The initial sizes of magnitude and colour cells were 
($\Delta\,(H-K_{\rm s})$,\,$\Delta\,H$)=(0.2,\,0.5).   
See \cite{mcs10} for more details on the decontamination 
method. 

\subsubsection{Application}
\begin{figure}
	\includegraphics[width=0.48\columnwidth]{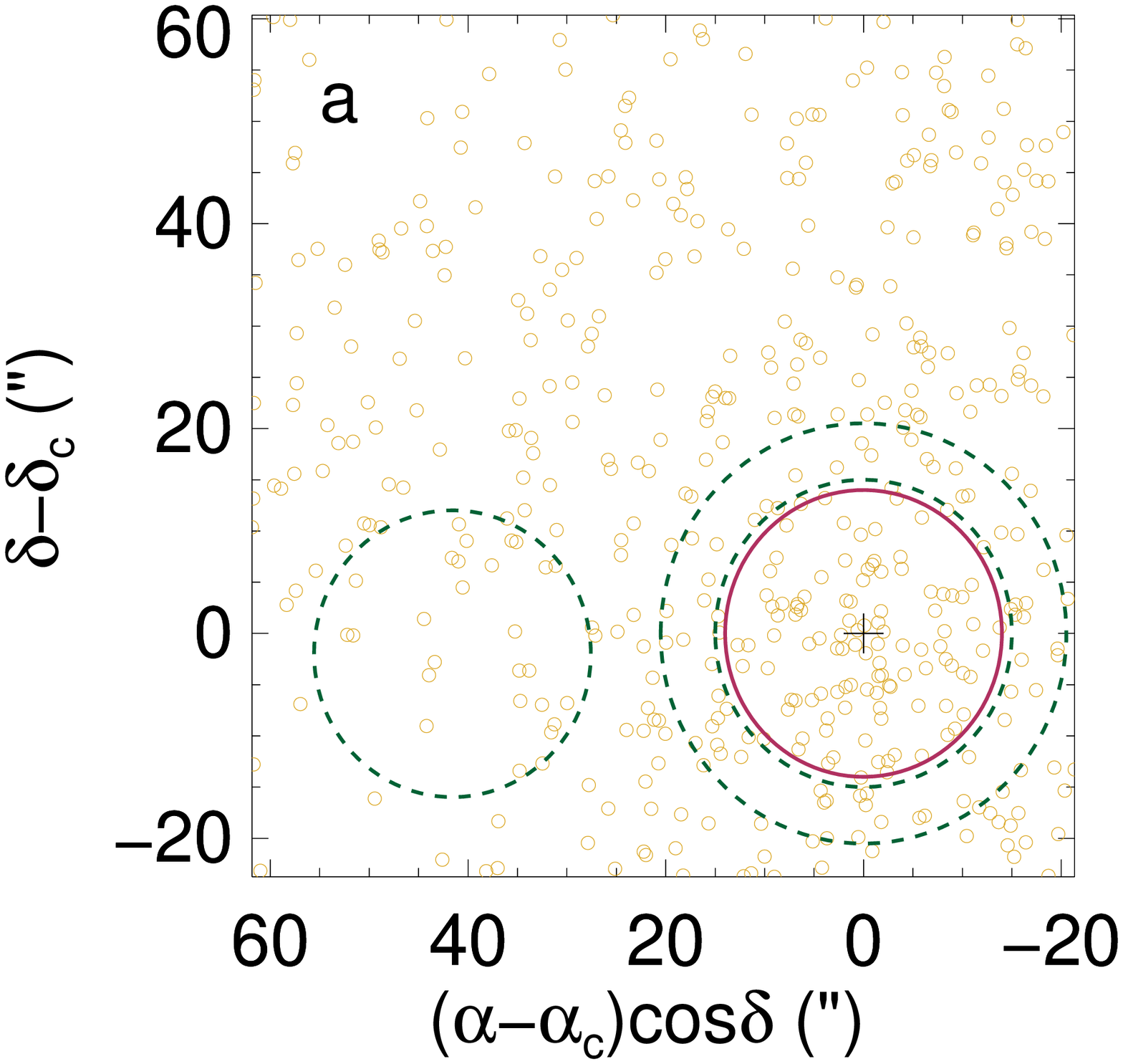}
	\includegraphics[width=0.48\columnwidth]{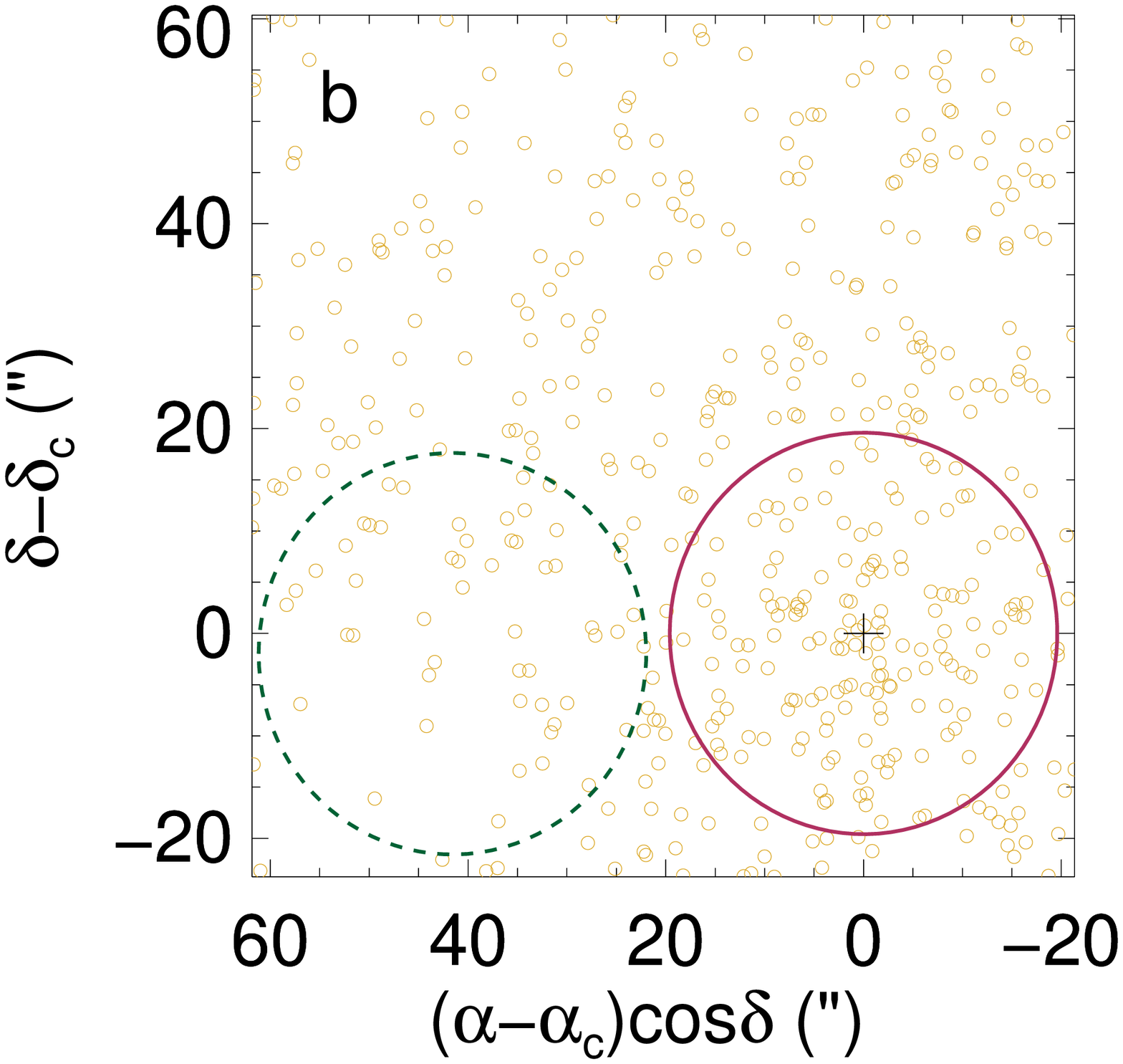}
\caption{(a) The positions of the two stellar fields (dashed lines) used to 
decontaminate the CMD of the cluster region (continuous line). All 
regions have the same area. The star positions (yellow circles) are 
indicated for stars with $H-K_{\rm s} > 0.8$ and $K_{\rm s} < 18$, for contrast
purposes. (b) Same as panel (a), but for a larger field and cluster areas.}
    \label{fig:fields2}
\end{figure}

\begin{figure}
	\includegraphics[width=\columnwidth]{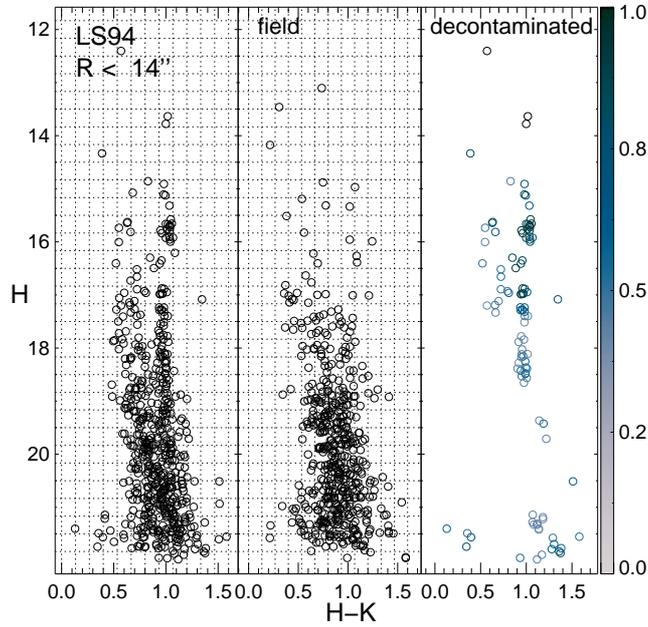}
\caption{Decontamination of LS\,94 CMD using as control field the annular area
surrounding the cluster. Left: the CMD for the 
spatial region of the cluster  defined by  $R<14$\,arcsec. Middle: the CMD of a
neighbouring  field covering the same area as the cluster region.
The grid shows one of the 729 configurations of cell positions and size
employed by the decontamination method. Right:
The CMD of the decontaminated sample
with stellar membership probabilities indicated by the colourbar.}
    \label{fig:decont_ring}
\end{figure}

\begin{figure}
	\includegraphics[width=\columnwidth]{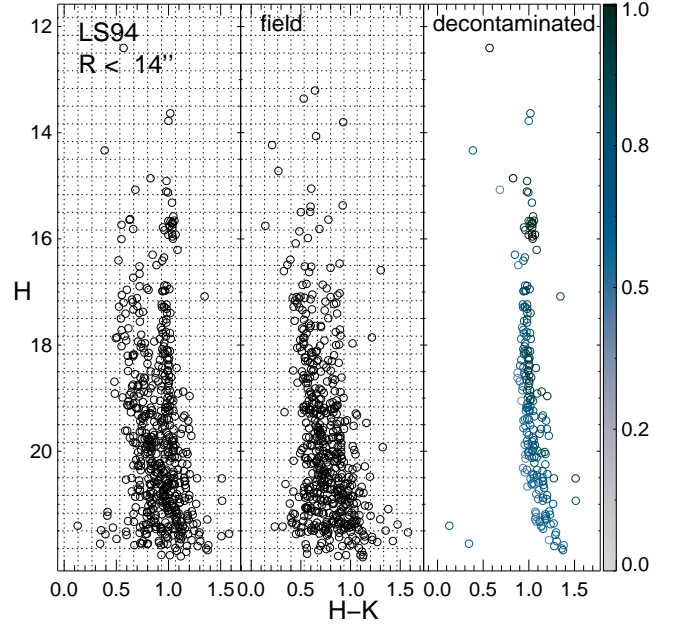}
\caption{Same as Fig.~\ref{fig:decont_ring} but using as control field the 
circular area adjacent to the cluster.}
    \label{fig:decont_circ}
\end{figure}

\begin{figure}
	\includegraphics[width=\columnwidth]{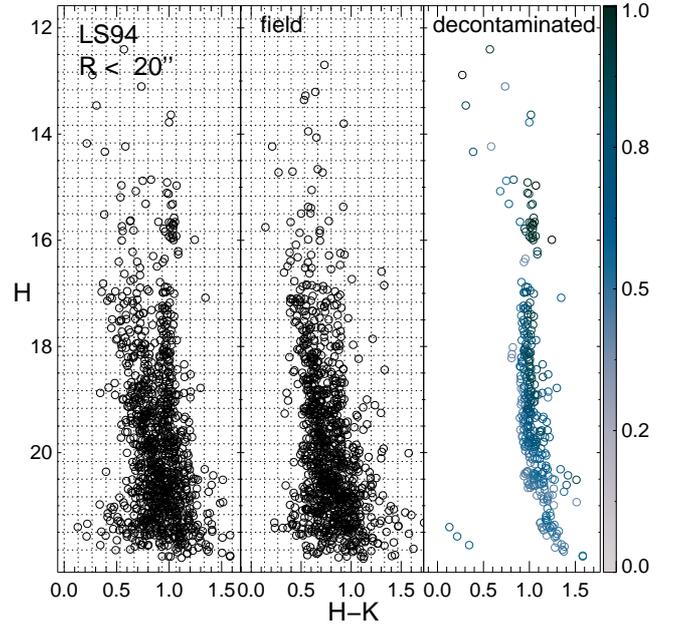}
\caption{Decontamination of LS\,94 CMD using the circular area
with $r<20$\,arcsec for both cluster and surrounding field. Symbols as in 
Fig.~\ref{fig:decont_ring}.}
\label{fig:decont_circ2} 
\end{figure}

Two control fields
(same area as the circular cluster region)
were employed to compare the results of decontaminating the central 
region of the cluster 
($r<14$\,arcsec): \textit{(i)}~ the region defined by a ring surrounding the cluster
from $15<r<20$\,arcsec and \textit{(ii)}~ the region displaced from the cluster
centre by 36\,arcsec along the same Galactic latitude, towards East mostly. 
Fig.~\ref{fig:fields2}a depicts these regions together with the cluster region
for stars with $H-K_{\rm s} > 0.8$ and $K_{\rm s} < 18$, just to enhance 
the contrast between cluster and field stars.
Figs.~\ref{fig:decont_ring} and \ref{fig:decont_circ} show the 
decontaminated CMDs using each control field, without any colour or 
magnitude filtering. The grid represents 
one cell configuration and the vertical colourbar reflects 
the membership probability assigned to each star.
It is clear that the annular field contains MS stars below the 
turnoff belonging to the cluster. The decontamination method eliminates most
stars with $H>18.5$. On the other hand, when the circular control field towards
East is considered, the low MS is retained by the 
decontamination method. Therefore, a better account of cluster members 
was pursued in which a larger circular area was considered to include
those MS stars. To do this, the decontamination method was applied to
fields of $r<20$\,arcsec, shown in Fig.~\ref{fig:fields2}b. As for the previous
analysis, the control field was displaced from LS\,94 centre towards East,
and at the same Galactic latitude. The result is presented in 
Fig.~\ref{fig:decont_circ2}.  

It is worth noticing that additional control fields
were tested toward other directions about 40\,arcsec from the cluster centre,
with similar results as the one obtained for the circular field. Since the
circular control field located as in Fig.\ref{fig:fields2}b has its centre 
at the 
same Galactic latitude as the cluster centre, it was assumed to give the best
representation of the field over the cluster area, minimising the disc stellar 
population gradient and, possibly, differential reddening. Indeed, the 
similar colour 
width of the stellar populations observed in both cluster and field 
CMDs of Fig.~\ref{fig:decont_circ2} 
makes us confident that the subjacent stellar population and the 
reddening are not very different for them.

\section{Fundamental parameters}
\label{sec:red_dist}

\subsection{Reddening}

The reddening was inferred from the decontaminated colour--colour
diagram (TCD)
$J-H$~vs~$H-K_{\rm s}$ and the intrinsic 2MASS near--infrared colours for 
dwarfs and giants covering a broad range of temperatures \citep{sl09}. 
Specifically, the intrinsic colours of
the RCG, where stars burn helium in their cores, were 
compared to the colours measured for the observed RCG. The RCG intrinsic 
averaged colors, ($J-H$)$_\circ=0.46\pm0.02$ and 
($H-K_{\rm s}$)$_\circ=0.09\pm0.03$,
are based on more than a hundred stars in six well--studied open clusters, 
and correspond to spectral type G8\,III \citep{sl09}. The average 
for the 19 RCG stars within 14\,arcsec of LS\,94 
yields $\langle J-H \rangle=1.94\pm0.11$ and 
$\langle H-K_{\rm s} \rangle=1.02\pm0.03$, which are values 
reddened, consequently, by E($J-H$)$=1.48\pm0.11$ and 
E($H-K_{\rm s}$)$=0.93\pm0.04$. 
Using the \citet{rl85} extintion law, the following 
quantities were derived:
$E(B-V)=4.59 \pm 0.23$, $A_V=14.18\pm 0.71$, $A_{K_{\rm s}}=1.59\pm 0.08$.

\begin{figure}
	\includegraphics[width=\columnwidth]{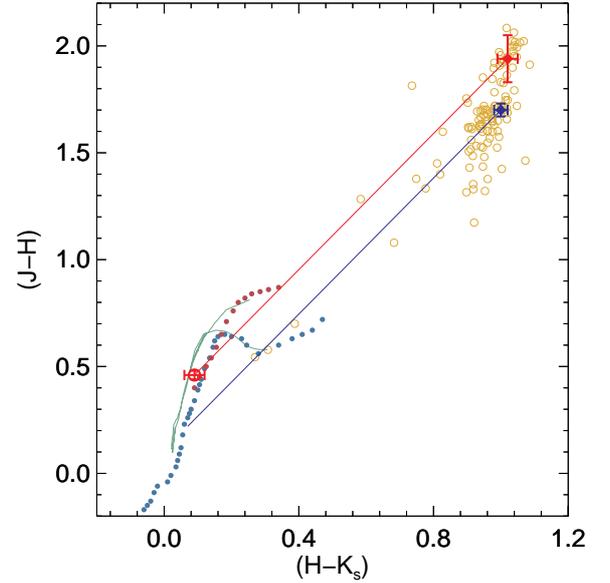}
\caption{Colour--colour diagram of LS\,94 members (yellow circles). 
Intrinsic colors of giants (red dots) and MS stars (blue dots) 
from \citet{sl09}. The red straight line connects intrinsic
and LS\,94 reddened positions of the RCG (red symbols 
with error bars). The blue straight line connects intrinsic and LS\,94 reddened
locus of the turnoff (blue symbol with error bars). The green continuous line
is the locus of an isochrone with $\log{t}=9.12$ and $Z=0.019$.}
    \label{fig:tcd}
\end{figure}

Fig.~\ref{fig:tcd} is the TCD
containing the stars considered members of LS\,94, the intrinsic sequences
of dwarfs and giants and the positions of the intrinsic RCG and of the 
LS\,94 reddened RCG. These positions are linked by a straight line 
representing the reddening towards the cluster inner core, where the RCG stars
are located. The locus of the best--fitting isochrone of age $\log{t}=9.12$ and 
metallicity $Z=0.019$ (see Sect.~\ref{sec:age_met}) is also shown there.

The position of the cluster turnoff is about
$\langle H \rangle=16.90\pm0.02$, $\langle H-K_{\rm s} \rangle=1.00\pm0.02$, 
$\langle J-H \rangle=1.70\pm0.03$. Using these
colors as starting point in the TCD, a new line was drawn keeping the 
same slope and extent as determined by the reddening calculated from the 
RCG. The line end point lies on the expected location of intrinsic colours
of dwarfs, which suggests that RCG giants and MS stars, with 
different spatial distributions over the cluster region, are affected by
nearly the same reddening. 
This analysis links the intrinsic colours of the cluster turnoff
with spectral type F8--G0\,V.

\begin{figure}
\centering
\includegraphics[width=\linewidth]{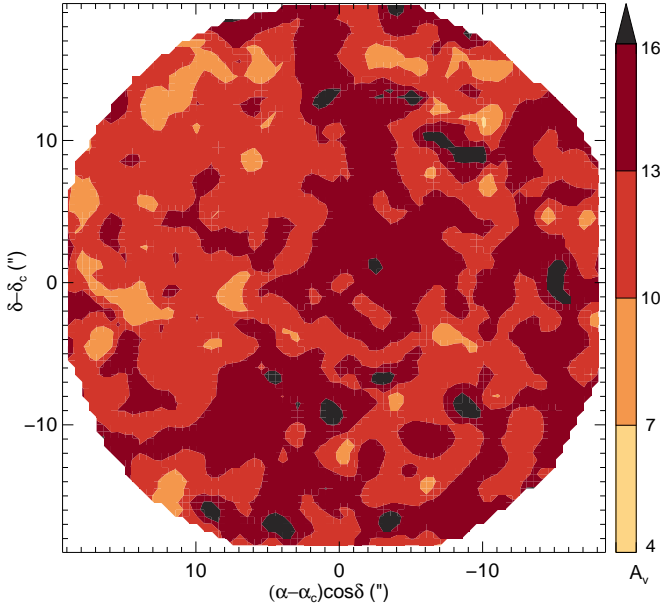}
\caption{Extinction map derived for the inner 20\,arcsec of LS\,94.}
\label{fig:ext}
\end{figure}

The spatial distribution of the extinction was also investigated by 
constructing an extinction map based on the complete sample of stars
within 20\,arcsec from the cluster centre. To this end, an extinction 
value was calculated
for each star by dereddening it along the reddening vector in the TCD, up 
to the MS locus defined by the $\log t$=9.12 ($Z=0.019$) isochrone. 
However, since only a small fraction of the sample possesses $J$ 
band magnitudes, we have repeated this procedure using the larger sample 
provided by the 
stars' $H-K_{\rm s}$ colour only and the isochrone mean intrinsic colour 
($H-K_{\rm s}$)$_\circ$=0.0827. Because the first 
extinction estimate was based on $JHK_s$ photometry benefiting from a broad
range of possible intrinsic colours, it was used to calibrate the less reliable
values obtained from employing exclusively $H$-$K_s$ colours, 
revealing that this latter approach 
overestimate extinction  by 1.0 mag on average 
\citep[for details on the method, see][]{mmj15}. 
These calibrated extinction values were interpolated into an uniform grid with 
a resolution of 0.5\,arcsec, the modal star separation in the field, and
finally smoothed by a 1.5\,arcsec width median kernel to build the final 
map shown in Fig.~\ref{fig:ext}. The map reveals
a complex pattern, dominated by a heavier extinction strip in the 
N-S direction, nearly perpendicular to the Galactic disc, presenting 
values between $13<A_V<16$. However, most of the region shows
lower extinction values. To derive the cluster's fundamental parameters we used
the extinction derived from the position of the RCG stars, as they are
clear members located in the cluster central region.

Confronting extinction (Fig.~\ref{fig:ext}) and stellar density 
(Fig.~\ref{fig:dens}) maps allowed us
to infer that the elongated shape formed by the stellar distribution inside the
cluster core cannot be produced by an artefact of enhanced extinction
around its E-W borders, since the central parts of the cluster have
higher extinction than its surroundings. 
Notwithstanding the existence of differential extinction in the region, 
the previous argument rules out the dust as responsible for the disturbed 
appearance of the cluster core.

\subsection{Distance}
The cluster distance was determined from the absolute magnitude of the RCG 
stars.
Their absolute magnitude is an efficient distance
indicator of a stellar system, especially in the near--infrared bands, 
where the uncertainties in the population age and metallicity are negligible 
compared to optical bands \citep{a00,gs02}.

\citet{vg07}, using 2MASS data for 24 open clusters with known distances,
obtain $M_{K_{\rm s}}$(RCG)$=-1.57\pm0.05$, arguing that this value is
reliable for distance determinations of clusters with metallicities between
$-0.5$ and $+0.4$\,dex and ages between approximately 300\,Myr and 8\,Gyr. 
Assuming this absolute magnitude and the value measured for the apparent
magnitude of the cluster RCG, i.e., 
$\langle K_{\rm s} \rangle$(RCG)$=K=14.67\pm0.24$, 
the true distance modulus was calculated adopting the extinction
$A_{K_{\rm s}}=1.59\pm 0.08$. The result is ($m-M$)$_\circ = 14.65\pm0.26$, 
which leads to a distance of $d=8.5\pm1.0$\,kpc for LS\,94.

\subsection{Location in the Milky Way}

The IAU--recommended value for the distance to the Galactic centre, 
$R_\circ =8.5$\,kpc,
has been revised by several authors using different methods.
A recent compilation (from studies between 1990 and 2012) aimed at 
estimating $R_\circ$, enabled \citet{gef13} to indicate as the most probable
value $R_\circ=8.20\pm0.35$\,kpc. This value was adopted to calculate the
cluster Galactocentric distance from its distance to the Sun 
derived above.
The result, $R=7.30\pm0.49$\,kpc, places the cluster $\sim 1$\,kpc inside 
the solar circle.

\begin{figure}
	\includegraphics[width=\columnwidth]{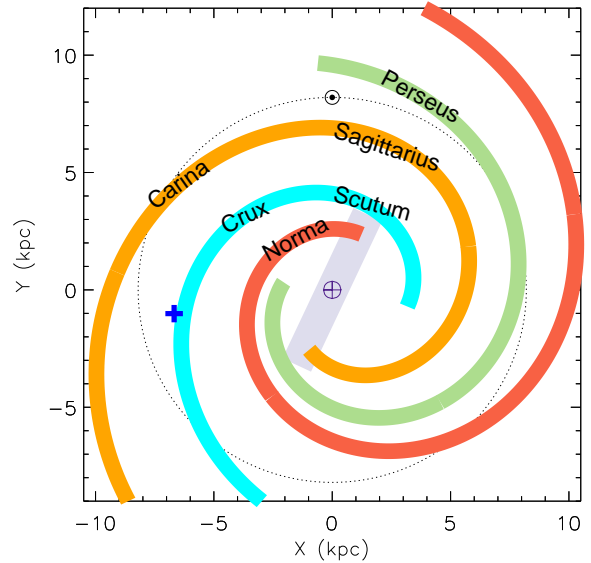}
\caption{Representation of the MW plane with known spiral arms marked.
The sun position and its orbit (dotted line) are shown together with the
the Galactic centre (crossed circle), the bar and the position of LS\,94
(blue cross).}
    \label{fig:spiral}
\end{figure}

Fig.~\ref{fig:spiral} depicts the MW plane with the main spiral
arms according to the representation by \citet{pmg10}, which is based on
\citet{v08}. The Galactic centre, the bar and the sun position are indicated, 
as well as the solar orbit (solar circle) and the position of LS\,94.
The cluster lies in the Crux arm, in the fourth quadrant, inside 
the solar circle.

\subsection{Age and metallicity}
\label{sec:age_met}

To derive age and metallicity, PARSEC isochrones version 1.2S 
\citep{bgm12,cbg15}
were fit to cluster CMDs decontaminated from field stars. Reddening and
distance modulus as derived in Sect.~\ref{sec:red_dist} were applied 
to the data before the isochone fitting.

\begin{figure}
	\includegraphics[width=\columnwidth]{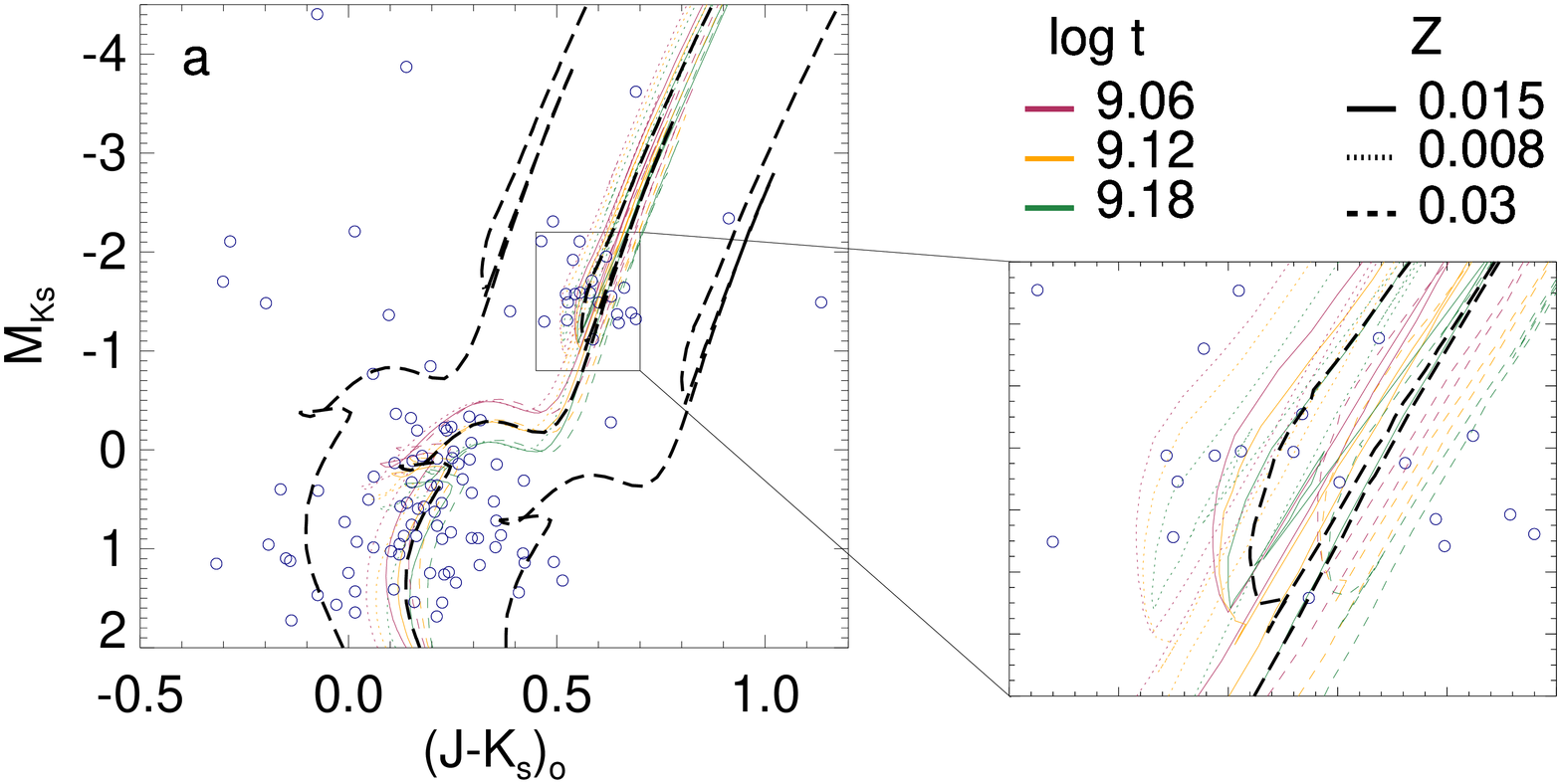}
	\includegraphics[width=\columnwidth]{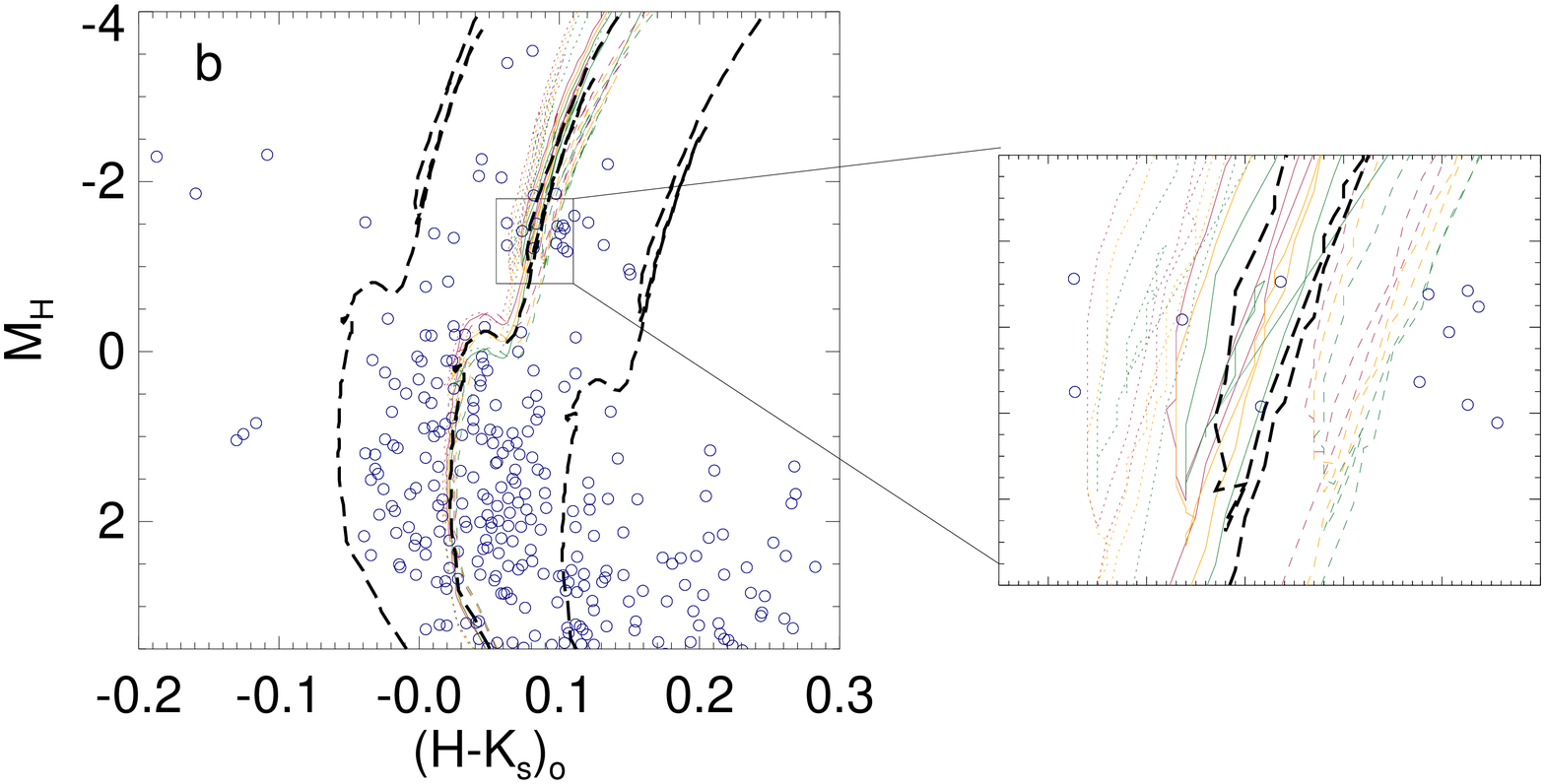}
\caption{ Isochrone match to the CMDs $M_{K_{\rm s}}$ vs ($J-K_{\rm s}$)$_\circ$
(panel a) and $M_H$ vs ($H-K_{\rm s}$)$_\circ$ (panel b). 
A set of isochrones covering ages $\log{t} = 9.06, 9.12, 9.18$ and 
metallicities $Z= 0.008, 0.015, 0.03$ is superimposed over the cluster 
CMDs. The best--fitting isochrone with $\log{t}=9.12$ and $Z=0.02$ 
is shown besided by the same isochrone shifted according to 
2 $\sigma$ uncertainties in extinction 
and distance (black long--dashed lines). The RCG region is zoomed to make 
clearer the isochrone 
differences and their match to the data.}
    \label{fig:isocfit}
\end{figure}

Fig.~\ref{fig:isocfit} shows the best--fitting isochrones to the intrinsic
CMDs $M_{K_{\rm s}}$ vs ($J-K_{\rm s}$)$_\circ$ and $M_H$ vs ($H-K_{\rm s}$)$_\circ$. 
{The region of RCG stars is zoomed to give a better sense of isochrone 
age and metallicity differences and how they match the data.}

Although the CMD composed of $J$ and $K_s$ bands is shallower than that 
involving $H$ and $K_s$ bands, the former provides a longer baseline, allowing 
a  more precise determination of age and metallicity. Also, to perform the 
reasonable isochrone fitting seen on the $M_H$ vs ($H-K_{\rm s}$)$_\circ$ CMD,
the extinction coefficient $A_H$ needed to be increased from 0.175\,$A_V$ to
0.178\,$A_V$, a value in agreement with extinction laws by \citet{ccm89}
and \citet{imb05}.
In conclusion, taking into account the range of isochrones that fit the
cluster intrinsic CMDs, the age and metallicity determined for LS\,94 are
$\log{t}=9.12\pm0.06$ and $Z=0.02\pm0.01$.

The effect of the uncertainties in the distance modulus and extinction is 
also presented in Fig.~\ref{fig:isocfit}, where the best--fitting 
isochrone with the age and metallicity
determined above is shown together with the same isochrone 
displaced by amounts given by the uncertainties (2 $\sigma$) in E($H-K_{\rm s}$)
(0.04), E($J-K_{\rm s}$) (0.12), $A_K$ (0.08), $A_H$ (0.12) and 
($m-M$)$_\circ$ (0.26). The data plotted cover the whole set of stars which
survived the decontamination method, although outliers that are clearly 
non--members due to their positions in the CMDs, were retained in 
Fig.~\ref{fig:isocfit}.

\section{Cluster structure}
\label{sec:struc}

\subsection{Radial density profile}

\begin{figure}
	\includegraphics[width=\columnwidth]{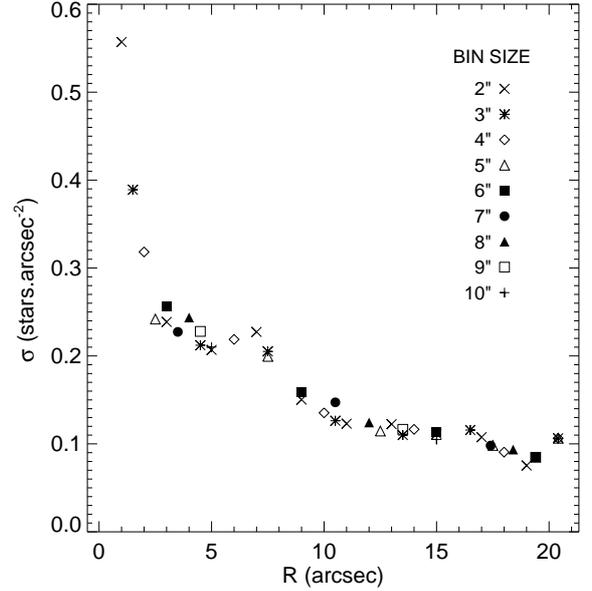}
\caption{LS\,94 radial density profile. Different bin sizes are indicated 
by different symbols.}
    \label{fig:rdp}
\end{figure}

The cluster structure was investigated by building its radial density profile
(RDP) using annuli of several widths to evaluate stellar densities from star 
counting. The annuli are 
centred in the derived cluster position (Sect. ~\ref{sec:map}).
Despite the cluster distortions evidenced by the density map 
(Sect.~\ref{sec:map}), the RDP shows a clear 
overdensity that falls significantly until 20\,arcsec, nearly at the border
of the frame (Fig.~\ref{fig:rdp}). 

The star counts were performed for data filtered in $K_{\rm s}$
($<18$) and $H-K_{\rm s}$ ($> 0.8$) to enhance cluster to field 
contrast. However, it is worth noticing
that these limits also mean that the RDP does not count
lower MS stars with $M_{K_{\rm s}}>1.76$ or masses below 1.56\,M$_\odot$
according to the derived distance, reddening and age 
(see Sect.~\ref{sec:red_dist}). 
Coupled with the small GSAOI FoV, the already mentioned extended 
population of lower MS stars, would make a determination of the cluster tidal
radius through the fitting of a three--parameter \citet{k62} model not
useful. An estimate of the cluster core radius is given in the next section,
but a discussion on its tidal radius is postponed to 
Sect.~\ref{sec:lm}, where the cluster mass is derived.

\subsection{Central surface density and core radius}

\begin{figure}
	\includegraphics[width=\columnwidth]{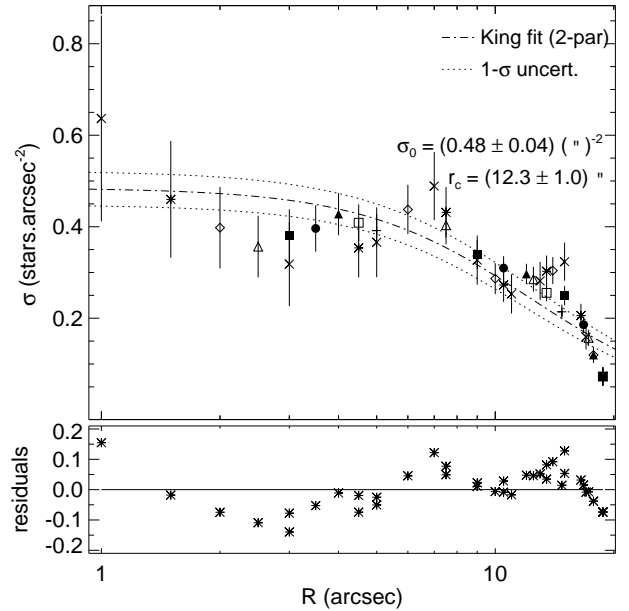}
\caption{LS\,94 RDP and the two--parameter King model fitting (dashed line) 
with envelope of 1\,$\sigma$ uncertainties (dotted lines). 
Symbols as in Fig.~\ref{fig:rdp}. 
The fitting residuals are also presented in the lower panel.}
    \label{fig:king2d}
\end{figure}

To estimate the central surface density and the core radius of LS\,94, the
CMD decontaminated sample (for $r<20$\,arcsec) up to the completness 
limit ($H<21.2$) was
subjected to a two--parameter \citet{k62} model fitting (Fig.~\ref{fig:king2d}). 
Because the selected data sample was already decontaminated, the stellar density
background is null. The fitting (Fig.~\ref{fig:king2d}) provides  
$\sigma_\circ=(0.48\pm0.04)$\,stars/arcsec$^2$ for the cluster central 
stellar density and $r_{\rm c}=(12.3\pm1.0)$\,arcsec for its core radius.
With the distance derived in Sect.~\ref{sec:red_dist}, the scale is 
1\,arcsec=0.0412\,pc and the converted values 
are $\sigma_\circ=283\pm24$\,stars/pc$^2$ and 
$r_{\rm c}= 0.51\pm0.04$\,pc, respectively.

Although the King model fitting was successful and useful to estimate
$\sigma_\circ$ and $r_{\rm c}$, the cluster 
RDP has wiggles and bumps, compatible with a system in an
advanced stage of evolution. Particularly, 
the central density is marginally described by the model. 
Taken directly from the observed RDP, the inner data point corresponds to
$\sigma_\circ=0.64\pm0.23$\,stars/arcsec$^2$ or  
$38\pm14\times 10$\,stars/pc$^2$.
The RDP central cusp is a known characteristic of evolved stellar
systems like globular clusters with colapsed cores \citep{tkd95}
and old open clusters \citep{mob08,b11}. 

Given the small FoV of GSAOI, a tidal radius was not fit, but obtained 
from the estimated total cluster mass, which was calculated by adding the 
observed stellar mass and that extrapolated to lower MS stars
according to a mass function (see Sect.~\ref{sec:rt}).

\subsection{Mass segregation}

\begin{figure}
	\includegraphics[width=\columnwidth]{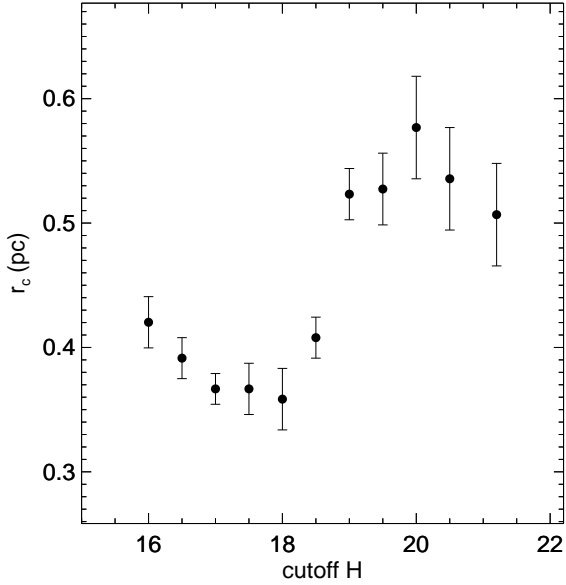}
\caption{The core radius variation as a function of the magnitude cutoff.
Stars brighter than the cutoff were included in the King model fitting
to derive the core radius.}
    \label{fig:rc_cut}
\end{figure}

Using the same decontaminated data as above, the two--parameter King
model fitting was performed for different $H$ magnitude cutoffs, i.e.,
in addition to the completeness limit $H=21.2$, the cutoff was progressively 
decreased from
20.5 to 16, with intervals of 0.5 mag. Recalling that the 
turnoff is around $H=16.9$, if the cutoff is at $H=17$ then only
evolved stars will be included in the fitting.
The results are presented in Fig.~\ref{fig:rc_cut}, where the $r_{\rm c}$
obtained from the profile fitting is plotted as a function of the
magnitude cutoff.

The cluster core radius increases whenever the magnitude cutoff 
includes more MS stars. It appears to be a transition around $H\sim 18.7$
where brighter stars are centrally concentrated ($r_{\rm c}\sim 0.38$\,pc)
while fainter stars are less concentrated ($r_{\rm c}\sim 0.55$\,pc).
Assuming that the overall core radius up to the completeness 
limit (Fig.~\ref{fig:king2d}) 
is a fiducial cluster core radius ($r_{\rm c}\sim 0.51\pm 0.04$\,pc), 
then it indicates that the mass segregation occurs essentially inside 
the cluster core. 

Beyond the population gradient shown in Fig.~\ref{fig:HK_H}, the 
structural parameters gave additional information about
the cluster dynamical state. In the next Section,
a quantitave analysis of the stellar population distribution is given,
also corroborating with these results.

\section{Cluster integrated properties and tidal radius}
\label{sec:lm}

\subsection{Mass}
\begin{figure}
	\includegraphics[width=\columnwidth]{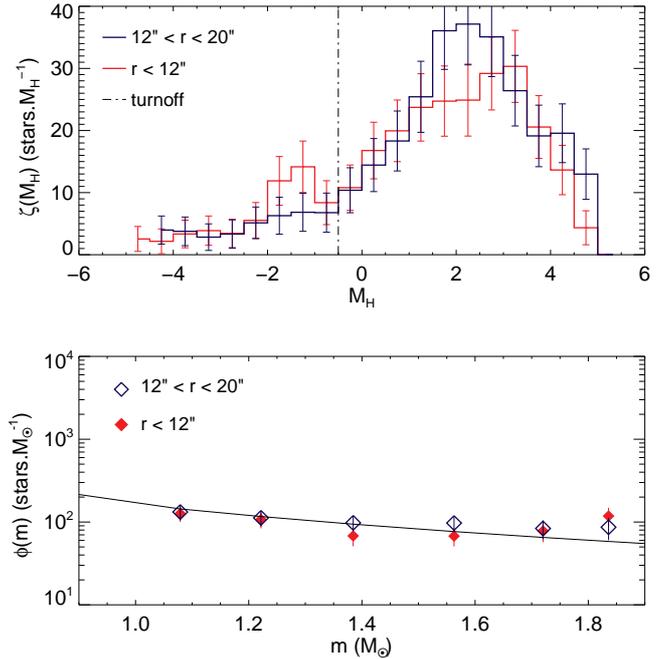}
\caption{The luminosity function (top) and the mass function (bottom)
of LS\,94 from the decontaminated sample for two regions centred in the 
cluster. The fitting to the mass function yielded 0.7 for its slope and
is represented by the continuous line. The uncertainties indicated by
error bars come from Poisson statistics applied to the star counts.}
    \label{fig:lf_mf}
\end{figure}

The cluster mass inside its core ($r_{\rm c}<12$\,arcsec)
was derived from the observed $M_H$ luminosity function (LF) converted 
to a mass function (MF) with the aid of the best--fitting isochrone
mass--luminosity relation. The LF for the core region is compared
with that of an external region ($r_{\rm c}<r<20$\,arcsec) 
in Fig.~\ref{fig:lf_mf}.
The LFs include all stars from the decontaminated sample. 
The outer region LF contains more stars because its area is larger than
that of the inner region and it may also be affected more significantly
by differential extinction.  
The range of stellar masses sampled by the LF ($-3.5<M_H<4.7$) is 
$2.13>m$(M$_\odot$)$>0.71$ with the turnoff ($M_H=-0.27$) mass 
at $\sim$1.96\,M$_\odot$. To model the stellar
mass distribution where it is unseen or incomplete, a power--law MF
was employed, namely $A=m^{-(1+x)}$, where $A$ is a
normalization constant and $x$ the slope. 

Although the completeness limit for the whole FoV in the $H$ band (21.2) 
reaches 
$M_H =4.02$ or $m=0.88$\,M$_\odot$, the MF normalization was chosen 
at the LF peak ($M_H\sim 3.5$ and $m=1.01$\,M$_\odot$), which is a better
constraint for the cluster (more crowded) region. 
Considering the core region, the total mass for stars more massive than 
1.01\,M$_\odot$ was estimated in $220\pm21$\,M$_\odot$. Between $1.01$ and 
$1.89$\,M$_\odot$, the distribution of MS stars was fit by the power--law giving
$x=0.7$ for its slope (Fig.~\ref{fig:lf_mf}), which served to 
extrapolate the MF down to 0.5\,M$_\odot$.
From there to the H--burning mass limit (0.08\,M$_\odot$) it was assumed that 
the slope flattens to $x=0.3$, in accord with a \citet{k01} MF. 
Integration of 
the MF yields the partial masses $91\pm6$\,M$_\odot$ between $1.01$ and 
$0.5$\,M$_\odot$ and $130\pm57$\,M$_\odot$ between 0.5 and 0.08\,M$_\odot$.
Therefore, the total mass in the core is the sum of the three quoted values,
i.e., $441\pm61$\,M$_\odot$.

\citet{bb05} analysed a sample of open clusters of different 
ages and concluded that there is a strong correlation between the
core mass and the overall mass, regardless of how populous is the 
cluster (see their fig. 9d). From our estimate for the core mass
and their results, it leads to the overall cluster mass of $6\pm1$
times bigger than that of the core, that is
$\mathcal{M}=(2.65\pm0.57)\times 10^3$\,M$_\odot$. The number of stars
associated with this mass is $\approx $3600.

\subsection{Tidal radius}
\label{sec:rt}

The cluster tidal radius is poorly constrained by the small region
sampled by GSAOI but it can be estimated using the cluster total mass
and the MW rotation curve parameters. According to 
\citet[][and references therein]{xz13}, the
MW rotation curve circular velocity at the Galactocentric distance 
of LS\,94 ($R=7.30\pm0.49$\,kpc) is $V_{\rm c}=250\pm30$\,km/s. 
Consequently, the Galaxy mass inside the cluster orbit results
$\mathcal{M}_g=1.00\pm0.07 \times 10^{11}$\,M$_\odot$.

The tidal radius of an open cluster with disc kinematics \citep{k62} can be 
estimated by

\begin{equation}
\label{tidal1}
r_{\rm t} = \left( \frac{G\mathcal{M}}{4\mathcal{A}(\mathcal{A}-\mathcal{B})} \right)^{1/3}
\end{equation}

where $\mathcal{A}=14.82\pm0.84$\,km\,s$^{-1}$\,kpc$^{-1}$ and 
$\mathcal{B}=-12.37\pm0.64$\,km\,s$^{-1}$\,kpc$^{-1}$ \citep{fw97} are  
the \citet{o27} constants.
Inserting into this equation the cluster mass $\mathcal{M}$ 
leads to $r_{\rm t}= 19.2\pm1.4$\,pc.

Another estimate for LS\,94 tidal radius may be obtained from the 
expression \citep{k62}:

\begin{equation}
\label{tidal2}
r_{\rm t} = R \left( \frac{\mathcal{M}}{3\mathcal{M}_g} \right)^{1/3}
\end{equation}
 
In this case, $r_{\rm t}=15.1\pm1.5$\,pc was obtained. 
Both equations \ref{tidal1} and \ref{tidal2} are equivalent
but give independent tidal radius estimates since they rely on different
sets of observational parameters. Averaging both values
provides our final estimate: $r_{\rm t}=17.2\pm2.1$\,pc.
Combining the derived cluster radii $r_{\rm t}$ and $r_{\rm c}$, the concentration
parameter ($c=\log{r_t/r_c}$) follows, $c=1.53\pm0.06$.
All these structural parameters estimates should be taken as approximations
considering that LS\,94, as already pointed out, is dynamically evolved
undergoing mass segregation and the models were designed to 
describe massive symmetric systems in dynamic equilibrium.

The relaxation time ($t_r$) of a stellar system with $N$ stars can be 
defined as $t_r=\frac{N}{8\ln{N}} t_{cr}$, where $t_{cr}=r/\sigma_V$ is the 
time--scale for a star to cross a distance $r$ with velocity $\sigma_V$
\citep{bt87}. The time--scale in which the cluster tends to kinetic energy
equipartition, transfering massive stars to its core and low mass stars 
to its corona is what $t_r$ measures. To calculate $t_r$ for LS\,94, a 
typical value of $\sigma_V \approx 3$\,km/s found for open clusters 
\citep{bm98} was
used together with $r_{\rm t}$ and $r_{\rm c}$ obtained above and 
also the respective number
of stars $N_t \approx 3600$ inside 
$r_{\rm t}$ and $N_c \approx 600$ inside $r_{\rm c}$.
The result is $t_r\approx 5.5$\,Myr for the cluster core and 
$t_r\approx 300$\,Myr for the whole cluster. Compared to the cluster
age, 1.3\,Gyr, the much shorter $t_r$ indicates that the cluster had
time to reach an advanced stage of relaxation (faster in the core)
compatible with the mass segregation observed.

\subsection{Luminosity}

\begin{figure}
	\includegraphics[width=\columnwidth]{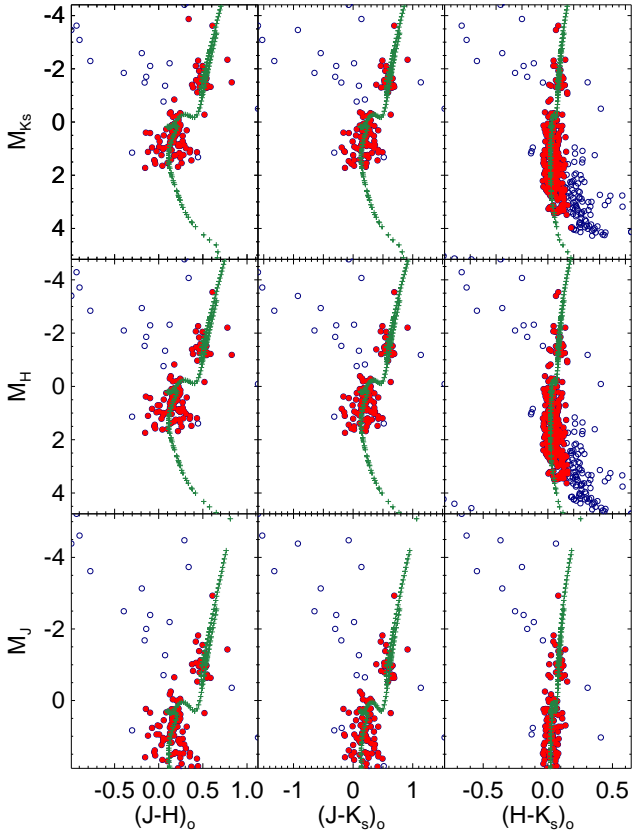}
\caption{Decontaminated CMDs with all possible combinations of 
magnitudes and colours.
Red filled circles indicate the subsample of stars that lie not farther
from the best--fitting isochrone than 3 $\sigma$ of the uncertainties 
in extinction. Blue open circles represent stars excluded from the
calculation of the integrated magnitude. The locus of the 
best--fitting isochrone is marked by green crosses.}
    \label{fig:cmdsel}
\end{figure}

The $J$, $H$ and $K_{\rm s}$ integrated absolute magnitudes 
were calculated by adding up the brightness of
selected member stars in the CMD.
The selected subsample of members was chosen according to a 
colour filter based on the best--fitting isochrone. Members that 
are farther from the isochrone than 3\,$\sigma$ of the uncertainty
in extinction are excluded from the calculation of integrated light.
The subsample was defined independently for
three CMDs of each magnitude as ordinate, e.g., the $J$ band was combined with
every possible colour $J-H$, $J-K_{\rm s}$ and $H-K_{\rm s}$. 
Fig.~\ref{fig:cmdsel}
shows the nine combinations of magnitudes and colours with
red symbols identifying the selected stars used to compute
the integrated magnitudes and green crosses marking the 
best--fitting isochrone ($\log{t}=9.12$ and $Z=0.02$) locus.

Therefore, for each row of Fig.~\ref{fig:cmdsel} there are three values of 
integrated magnitude obtained for the same band, which rely on independent
datasets. Whenever the J band is involved, the CMD reaches a shallower 
magnitude limit. To account for the low MS in all CMDs, the observed 
LF was extrapolated from the completeness limit according to the 
MF used to determine the cluster mass. The sum of the flux 
of individual stars added to the flux extrapolated using the MF
gives the integrated magnitude. There are two important points to
raise in this context: \textit{(i)} 
the integrated magnitudes are dominated by
bright stars, which are subject to stochastic effects that in turn 
induce fluctuations
in the integrated light \citep[][and references therein]{c13}, 
especially in the near--infrared for intermediate--age
to old clusters, which means that clusters with significantly different
integrated magnitudes may underlie stellar populations with similar
age and metallicity \citep{sf97}; \textit{(ii)} the 
decontamination method does not perform efficiently in CMD regions where the 
number of stars is small, as is in the present case for bright stars,
i.e.,  discriminating between bright field and member stars is hampered by
the small number statistics. This is why a colour filter helps 
to better constrain the cluster population.
Keeping these points in mind,  
the final integrated magnitudes are
$M_J= -5.30\pm0.06$, $M_H=-5.70\pm0.17$ and $M_{K_{\rm s}}=-5.82\pm0.16$.
These are, indeed, lower limits for the cluster integrated light since
they were estimated for stars within $r<20$\,arcsec $=0.82$\,pc, 
about 1.6 times the cluster core radius. However, bright member stars are not
expected to be found beyond the cluster core given its advanced stage of
evolution.

To check the consistency between the integrated magnitudes and mass 
estimates, simple stellar population models were built from the
$\log{t}=9.12$ and $Z=0.019$ isochrone with stars distributed as 
prescribed by the same MF employed above. The estimated integrated 
magnitudes were interpolated in the models to get the mass, resulting
$3.29\pm0.77\times 10^3$\,M$_\odot$, which is in agreement, within the 
uncertainties, with the total mass estimated for the cluster. 

\section{Discussion}
\label{sec:dis}

The final parameters adopted for LS\,94 are summarized 
in Table~\ref{tab:param}.

The differential reddening across the cluster area cannot reproduce 
the observed elongated stellar distribution of the cluster core,
as argued in Sec.~\ref{sec:red_dist}. Therefore, the distortions showing 
up in the surface density map reveal a stellar system disturbed by 
gravitational interactions with the Galactic disc material.
With 1.3\,Gyr and located inside the solar
circle, at $R=7.3$\,kpc from the Galactic centre, the cluster should have
completed many orbits, loosing stars to the Galactic field by
a combination of internal stellar and dynamical evolution with 
external processes
like disc shocking, molecular cloud encounters and gravitational 
stresses from spiral arms \citep{s87,gpb06}. All these 
external mechanisms impart 
cumulative tidal effects to the cluster after
many passages on its way throughout the disc.
The elongated shape of LS\,94 core, with semimayor axis oriented 
perpendicular to the Galactic disc is expected if the main 
mechanism actuating in the present time on the cluster is disc shocking
\citep{blg01,dmc15}. To clarify this issue, 
further investigation of the cluster 
kinematics would be needed.

The RDP presents a cusp in the very centre, characteristic
of evolved clusters. Indeed, mass segregation in the cluster core
seems to be  occurring, with most of the RCG stars (14) concentrating 
within $r<0.4$\,pc (10\,arcsec) and MS stars distributing themselves 
more evenly by the cluster core and outskirts. This behaviour is also
detected as an increase of the core radius with the magnitude level 
cutoff defining the star sample. Stars more massive than 
$m\sim 1.65$\,M$_\odot$ ($H=18.7$) are spatially distributed accordingly
to a King model with $r_{\rm c}\sim 0.38$\,pc, while stars less
massive than this value are better represented by $r_{\rm c}\sim 0.55$\,pc.
In addition, the slope of the cluster MF $x=0.7$ is flatter than that of the 
\citet{k01} MF for masses higher than 0.5\,M$_\odot$, also suggesting 
mass segregation.

\cite{cgc14} report 12 clusters older than 1\,Gyr and 
closer to the Galactic centre than the solar circle. Although
there are many more clusters within this selection criterion,
they were chosen for their well--determined ages and distances.
In this context, LS\,94 parameters fits among those of this sample, 
including the higher than solar metallicity (recalling that $Z=0.015$ 
is the sun metallicity for PARSEC isochrones).

Concerning structural parameters, LS\,94 is not as loose as those
investigated by \citet{bb05}, which studied eleven open clusters spanning
broad age and mass ranges. The six clusters with ages above 1\,Gyr
(three of them inside the solar circle) have concentration parameter
around $c\sim 1.0$, lower than that calculated for 
LS\,94 ($c\sim 1.5$). Comparing with Galactic globular clusters
\citep[][(2010 edition)]{h96}, the LS\,94 concentration parameter
is within the average: 26 per cent of the globulars out of 141 
with structure information have $c=1.5\pm0.2$.

\begin{table}
	\centering
	\caption{Parameters determined for LS\,94.}
	\label{tab:param}
	\begin{tabular}{lr}
		\hline
                $E(B-V)$  & $4.59 \pm 0.23$\\
		$A_V$ & $14.18\pm 0.71$\\
		$A_{K_{\rm s}}$ & $1.59\pm 0.08$\\
		$(m-M)_\circ$   & $14.65\pm 0.26$ \\
	        $d$ (kpc) & $8.5\pm 1.0$\\
	        $age$ (Gyr) & $1.3\pm 0.2$\\
		$Z$ & $0.02\pm0.01$\\
        	$R$ (kpc) & $7.30\pm 0.49$\\
		$\sigma_\circ$ (10\,stars/pc$^2$) & $38\pm14$ \\
		$r_{\rm c}$ (pc) & $0.51\pm0.04$\\
		$r_{\rm t}$ (pc) & $17.2\pm2.1$\\
		$c$  & $1.53\pm0.06$\\
                $t_{\rm r}$\,core  (Myr)&$\approx 5.5$\\ 
                $t_{\rm r}$\,overall (Myr)&$\approx 300$\\
	        $\mathcal{M}$ ($10^3$\,M$_\odot$) & $2.65\pm0.57$  \\
 		$M_{K_{\rm s}}$   & $-5.82\pm0.16$ \\
		\hline
	\end{tabular}
\end{table}

\section{Summary and concluding remarks}
\label{sec:conc}

Physical properties were derived for the candidate open cluster La Serena\,94,
recently unveiled by the VVV collaboration. The object's position 
is in the Galactic midplane under the influence of severe extinction 
towards the Crux spiral arm.
Deep photometry in $JHK_{\rm s}$--bands from GeMs/GSAOI was employed to 
characterize the object. The projected stellar density distribution of the 
region, conveyed into a 2D map, provided information on the location of 
the cluster centre and its overall structure. An analysis of the stellar 
population radial variation from the determined centre showed direct 
evidence of mass segregation with RCG stars centrally concentrated,
while MS stars spread farther into the cluster outskirsts.
Decontaminated $JHK_{\rm s}$ diagrams were built reaching stars with 
about 5 mag below the cluster turnoff in $H$. The locus of RCG stars 
in the TCD, together with an extinction law, 
was used to obtain an average extinction of $A_V=14.18\pm0.71$. The same
stars were considered as standard--candles to derive the cluster heliocentric 
distance, $d=8.5\pm 1.0$\,kpc.
Isochrones were matched to the member stars locii in CMDs to derive 
age ($\log{t}=9.12\pm 0.06$)
and metallicity ($Z=0.02\pm0.01$).
The cluster structure was investigated further by fitting King models
to its RDP, in spite of a central cusp and ragged appearance.
An overall core radius of $r_{\rm c}=0.51\pm 0.04$\,pc was obtained.
King models fittings to magnitude limited star samples evidenced mass
segregation as well, as the core radius shrinks from fainter 
to brigther stellar samples.
The cluster core mass was derived by adding up the visible stellar mass to an
extrapolated MF built from the LF in the $H$--band. 
A correction to this mass leads to $\mathcal{M}=(2.65\pm0.57)\times 10^3$\,M$_\odot$ for the
cluster total mass. The $JHK_{\rm s}$ integrated magnitudes were computed by summing up the star fluxes, resulting 
$M_J= -5.30\pm0.06$, $M_H=-5.70\pm0.17$ and $M_{K_{\rm s}}=-5.82\pm0.16$.
Consistency between mass and magnitude estimates 
were checked by comparing them with those of synthetic stellar populations
of same age and metallicity as those of the cluster. With the cluster mass 
determined, an estimate of the tidal radius was possible: 
$r_{\rm t}=17.2\pm2.1$\,pc.

The fundamental parameters of LS\,94 confirm that it is an old open 
cluster located in the Crux spiral arm
towards the fourth Galactic quadrant and distant $7.30\pm 0.49$\,kpc from
the Galactic centre. From our perspective, the cluster light propagates
roughly 3\,kpc through the Crux arm and another 5.5\,kpc through the
Galactic disc before reaching us.
The cluster age and distorted structure already 
suggested that it is a dinamically 
evolved stellar system. Also, its position inside the solar circle 
is expected to speed up the cluster
dynamical evolution in consequence of stronger tidal effects \citep{blg01,bb05}.
Further analyses confirmed that, indeed, with
an overall relaxation time 4 times shorter than its age and clear
evidences of mass segregation, the cluster is at the final stages of
evolution before the remnant phase when most of the stars are
lost into the Galactic disc \citep[][and references therein]{pkb11}.
This conclusion was also confirmed by the structural
analysis of the cluster RDP, which showed a transition of the
core radius for stars in different mass intervals, and 
the stellar mass distribution, which revealed a shallower MF slope
compared with a Kroupa MF.

Continuing efforts to uncover distant clusters and derive their properties
are needed to fill the gap which reveal our Galaxy structure, as open clusters
are one of its tracers \citep{dl05,fm08}. In particular, observations 
of obscured, distant star clusters with 8--m class telescopes and 
sensitive near--infrared instruments incorporating AO, 
like GeMs/GSAOI, will contribute to increase the sample of well--studied 
clusters towards the third and fourth Galactic quadrants, where few 
systems have been characterized.

\section*{Acknowledgements}
We thank the anonymous referee for helping to improve this paper.
ARL thanks partial financial supported by the DIULS Regular project PR15143.
We thank Gemini Observatory commissioning team (technicians, engineers and science staff) for their efforts to make a reality GeMS/GSAOI and  collect the wonderful data presented in this paper.









\bsp	
\label{lastpage}
\end{document}